\documentclass[matbio]{svjour} 

\usepackage{times,amscd,amsmath,amssymb,amsbsy,epsfig}

\newcommand{\picturesAB}[3]{
\centerline{\raise #3 \hbox{(a)}
\psfig{file=#1,height=#3}
\hspace*{.2in}
\raise #3 \hbox{(b)}
\psfig{file=#2,height=#3}
}}
\newcommand{\picturesABC}[4]{
\centerline{\raise #4 \hbox{(a)}
\psfig{file=#1,height=#4}
\hspace*{.2in}
\raise #4 \hbox{(b)}
\psfig{file=#2,height=#4}
\hspace*{.2in}
\raise #4\hbox{(c)}
\psfig{file=#3,height=#4}
}}
\def\cf{{\em cf.~\relax}}%
\def\etal{{\em et \thinspace al.~\relax}}%
\def\ie{{\em i.\thinspace e.\relax}}%
\def\norm#1{\parallel#1\parallel}%

\newcommand{\boldU}{{\bf U}}
\newcommand{\boldB}{{\bf B}}

\newcommand{\boldD}{{\bf D}}
\newcommand{\boldJ}{{\bf J}}
\newcommand{\boldJone}{{\bf J}_1}

\newcommand{\cL}{{\cal L}}
\newcommand{\cT}{{\cal T}}

\newcommand{\dd}{\mbox{d}}

\newcommand{\dt}{\mbox{d}t}

\newcommand{\dz}{\mbox{d}z}
\newcommand{\dphi}{\mbox{d}\phi}
\newcommand{\boldj}{\mathbf{j}}
\newcommand{\bolde}{\mathbf{e}}
\newcommand{\boldn}{\mathbf{n}}

\newcommand{\boldq}{\mathbf{q}}
\newcommand{\boldu}{\mathbf{u}}
\newcommand{\boldx}{\mathbf{x}}
\newcommand{\boldy}{\mathbf{y}}
\newcommand{\boldz}{\mathbf{z}}
\newcommand{\boldf}{\mathbf{f}}
\newcommand{\boldv}{\mathbf{v}}
\newcommand{\boldA}{\mathbf{A}}
\newcommand{\boldF}{\mathbf{F}}
\newcommand{\boldG}{\mathbf{G}}
\newcommand{\boldI}{\mathbf{I}}
\newcommand{\boldX}{\mathbf{X}}
\newcommand{\boldY}{\mathbf{Y}}
\newcommand{\calD}{\cal{D}}
\newcommand{\calB}{\cal{B}}
\newcommand{\boldV}{\mathbf{V}}
\newcommand{\boldg}{\mathbf{g}}
\newcommand{\boldtau}{\boldsymbol{\tau}}
\newcommand{\boldpsi}{\boldsymbol{\psi}}
\newcommand{\boldLam}{\boldsymbol{\Lambda}}
\newcommand{\boldLamone}{\boldsymbol{\Lambda_1}}
\newcommand{\dboldq}{\mbox{d}\mathbf{q}}
\newcommand{\dboldx}{\mbox{d}\mathbf{x}}
\newcommand{\dboldv}{\mbox{d}\mathbf{v}}
\newcommand{\dboldy}{\mbox{d}\mathbf{y}}
\newcommand{\er}{\mathbb{R}}
\newcommand{\et}{{\bf e}}

\newcommand{\normin}[2]{\bfvert #1 \bfvert\hbox{\raise -0.5mm \hbox{$_{#2}$}}}

\newcommand{\tS}{{\tilde{S} }}

\newcommand{\MpicturesAB}[3]{
\centerline{\raise #3 \hbox{(a)}
\psfig{file=#1,height=#3}
\hspace*{.05in}
\raise #3 \hbox{(b)}
\psfig{file=#2,height=#3}
}}
\newcommand{\MpicturesABC}[4]{
\centerline{\raise #4 \hbox{(a)}
\psfig{file=#1,height=#4}
\hspace*{.05in}
\raise #4 \hbox{(b)}
\psfig{file=#2,height=#4}
\hspace*{.05in}
\raise #4\hbox{(c)}
\psfig{file=#3,height=#4}
}}

\newcommand{\norminex}[3]{\bfvert 
#1 \bfvert^{#3}_{\hbox{\raise -0.5mm\hbox{$_{#2}$}}}}
\begin{document}

\title{Taxis Equations for Amoeboid Cells}
\author{Radek Erban \and Hans G. Othmer}
\institute{Radek Erban \at
Mathematical Institute, 
University of Oxford, 
24-29 St Giles', 
Oxford, OX1 3LB,
United Kingdom, 
\email{erban@maths.ox.ac.uk}
\and
Hans G. Othmer \at
School of Mathematics,
270A Vincent Hall,
University of Minnesota,
Minneapolis, MN 55455,
USA,
\email{othmer@math.umn.edu}
}

\date{\today}
\titlerunning{Taxis Equations for Amoeboid Cells}
\authorrunning{R. Erban and H. G. Othmer}
\maketitle

\noindent
{\bf Abstract} The classical macroscopic chemotaxis equations have
previously been derived from an individual-based description of the
tactic response of cells that use a ``run-and-tumble'' strategy in
response to environmental cues \cite{Erban:2004:ICB,Erban:2005:STS}. 
Here we derive macroscopic equations
for the more complex type of behavioral response characteristic of
crawling cells, which detect a signal, extract directional information
from a scalar concentration field, and change their motile behavior
accordingly. We present several models of increasing complexity for
which the derivation of population-level equations is possible, and we
show how experimentally-measured statistics can be obtained from the
transport equation formalism. We also show that amoeboid cells that do
not adapt to constant signals can still aggregate in steady gradients,
but not in response to periodic waves. This is in contrast to the case
of cells that use a ``run-and-tumble'' strategy, where adaptation is
essential.

\section{Introduction}

Motile organisms sense their environment and can respond to it by (i)
directed movement toward or away from a signal, which is called {\em
taxis}, (ii) by changing their speed of movement and/or frequency of
turning, which is called {\it kinesis}, or (iii) by a combination of
these. Usually these responses are both called taxes, and we adopt
this convention here.  Taxis involves three major components: (i) an
external signal, (ii) signal transduction machinery for transducing
the external signal into an internal signal, and (iii) internal
components that respond to the transduced signal and lead to changes
in the pattern of motility.  In order to move away from noxious
substances (repellents) or toward food sources (attractants) organisms
must extract directional information from an extracellular scalar
field, and there are two distinct strategies that are used to do
this. A simple paradigm will illustrate these.

Suppose that one is close enough to a bakery to detect the odors, but
cannot see the bakery. To find it, one strategy is to use sensors at
the end of each arm that measure the difference in the signal at the
current location and use the difference to decide on a
direction. Clearly humans do not use this strategy, but instead,
execute the ``bakery walk'', which is to take a sniff and judge the
signal intensity at the present location, take a step and another
sniff, compare the signals, and from the comparison decide on the next
step.

The first strategy is used by amoeboid cells (cells which move by
crawling through their environment), which have receptors on the cell
membrane and are large enough to detect typical differences in the
signal over their body length.  Small cells such as bacteria cannot
effectively make a ``two-point in space'' measurement over their body
length, and therefore they adopt the second strategy and measure the
temporal variation in the signal as they move through the external
field.  In either case, an important consideration in understanding
population-level behavior is whether or not the individual merely
detects the signal and responds to it, or whether the individual
alters it as well, for example by consuming it or by amplifying it so
as to relay the signal.  In the former case there is no feedback from
the local density of individuals to the external field, but when the
individual produces or degrades the signal, there is coupling between
the local density of individuals and the intensity of the signal. The
latter occurs, for example, when individuals move toward a signal from
neighboring cells and relay the signal as well, as in the aggregation
of the cellular slime mold {\em Dictyostelium discoideum} (Dd).

One of the best-characterized systems that adopts the ``bakery walk"
strategy is the flagellated bacterium \emph{E. coli}, for which the
signal transduction machinery is well characterized
\cite{Bourret:2002:MIP}.  \emph{E. coli} alternates between a more or
less linear motion called a run and a highly erratic motion called
tumbling, which produces little translocation but reorients the cell.
Run times are typically much longer than the tumbling time, and when
bacteria move in a favorable direction (\emph{i.e.}, either in the direction
of foodstuffs or away from noxious substances), the run times are
increased further.  Since these bacteria are too small to detect
spatial differences in the concentration of an attractant on the scale
of a cell length, they choose a new direction essentially at random at
the end of a tumble, although it has some bias in the direction of the
preceding run \cite{Berg:1972:CEC}.  The effect of alternating these
two modes of behavior and, in particular, of increasing the run length
when moving in a favorable direction, is that a bacterium executes a
three-dimensional random walk with drift in the favorable direction,
when observed on a sufficiently long time scale
\cite{Berg:1975:HBS,Koshland:1980:BCM}.  In addition, these bacteria
adapt to constant signal levels and in effect only alter the run
length in response to changes in extracellular signals.  Models for
signal transduction and adaptation in this system has been developed
\cite{Spiro:1997:MEA,Barkai:1997:RSB}, and a simplified version of the
first model has been incorporated into a population-level description
of behavior
\cite{Erban:2004:ICB,Erban:2005:STS}. The latter analysis shows how
parameters that characterize signal transduction and response in
individual cells are embedded in the macroscopic sensitivity $\chi$ in
the macroscopic chemotaxis equation described later. Having
the bacterial example in mind, we will call the ``bakery walk"
strategy as a ``run-and-tumble" strategy in what follows.

The directed motion of amoeboid cells (e.g. Dd or leukocytes), which
is crucial in embryonic development, wound repair, the immune response
to bacterial invasion, and tumor formation and metastasis, is much
more complicated than bacterial motion.  Cells detect extracellular
chemical and mechanical signals via membrane receptors, and these
trigger signal transduction cascades that produce intracellular
signals.  Small differences in the extracellular signal over the cell
are amplified into large end-to-end intracellular differences that
control the motile machinery of the cell and thereby determine the
spatial localization of contact sites with the substrate and the sites
of force-generation needed to produce directed motion
\cite{Parent:1999:CSD,Chung:2001:SPC}.  Movement of Dd and other
amoeboid cells involves at least four different stages
\cite{Mitchison:1996:ABC,Sheetz:1999:CMF}.  (1) Cells first extend
localized protrusions at the leading edge, which take the form of
lamellipodia, filopodia or pseudopodia.  (2) Not all protrusions are
persistent, in that they must anchor to the substrate or to another
cell in order for the remainder of the cell to follow
\cite{Soll:1995:UCU}. Protrusions are stabilized by formation of
adhesive complexes, which serve as sites for molecular signaling and
also transmit mechanical force to the substrate.  (3) Next, in
fibroblasts actomyosin filaments contract at the front of the cell and
pull the cell body toward the protrusion, whereas in Dd, contraction
is at the rear and the cytoplasm is squeezed forward.  (4) Finally
cells detach the adhesive contacts at the rear, allowing the tail of
the cell to follow the main cell body. In Dd the adhesive contacts are
relatively weak and the cells move rapidly ($\sim 20 \mu$m/min),
whereas in fibroblasts they are very strong and cells move slowly. The
coordination and control of this complex process of direction sensing,
amplification of spatial differences in the signal, assembly of the
motile machinery, and control of the attachment to the substratum
involves numerous molecules whose spatial distribution serves to
distinguish the front from the rear of the cell, and whose temporal
expression is tightly controlled.  In addition, Dd cells adapt to the
mean extracellular signal level \cite{Parent:1999:CSD}.

Dd is a widely-used model system for studying signal transduction,
chemotaxis, and cell motility.  Dd uses cAMP as a messenger for
signaling initiated by pacemaker cells to control cell movement in various
stages of development (reviewed in \cite{Othmer:1998:OCS}).  In the
absence of cAMP stimuli Dd cells extend pseudopods in more-or-less
random directions, although not strictly so since formation of a
pseudopod inhibits formation of another one nearby for some
time. Aggregation-competent cells respond to cAMP stimuli by
suppressing existing pseudopods and rounding up (the ``cringe
response''), which occurs within about 20 secs and lasts about 30 secs
\cite{Condeelis:1990:MAC}.  Under uniform elevation of the ambient
cAMP this is followed by extension of pseudopods in various
directions and an increase in the motility
\cite{Wessels:1992:BDA,Wessels:2000:IPR}.
However, one pseudopod usually dominates, even under uniform
stimulation. A localized application of cAMP elicits the ``cringe
response" followed by a localized extension of a pseudopod near the
point of application of the stimulus \cite{Swanson:1982:LSC}.  How the
cell determines the direction in which the signal is largest, and how
it organizes the motile machinery to polarize and move in that
direction, are major questions from both the experimental and
theoretical viewpoint.  Since cAMP receptors remain uniformly
distributed around the cell membrane during a tactic response,
receptor localization or aggregation is not part of the response
\cite{Jin:2000:LGP}.  Well-polarized cells are able to detect and
respond to chemoattractant gradients with as little as a 2\%
concentration difference between the anterior and posterior of the
cell \cite{Mato:1975:SIC}.  Directional changes of a shallow gradient
induce polarized cells to turn on a time scale of 2-3 seconds
\cite{Gerisch:1975:CCC}, whereas large changes lead to
large-scale disassembly of motile components and creation of a new
``leading edge'' directed toward the stimulus \cite{Gerisch:1982:CD}.
Polarity is labile in cells starved for short periods in that cells
can rapidly change their leading edge when the stimulus is moved
\cite{Swanson:1982:LSC}.

 There are a number of models for how cells extract directional
 information from the cAMP field. Fisher \etal \cite{Fisher:1989:QAC} suggest
 that directional information is obtained by the extension of
 pseudopods bearing cAMP receptors and that sensing the temporal
 change experienced by a receptor is equivalent to sensing the spatial
 gradient. However, Dd cells contain a cAMP-degrading enzyme on their
 surface, and it has been shown that as a result, the cAMP
 concentration increases in all directions normal to the cell surface
 \cite{Dallon:1998:CAC}. Furthermore, more recent experiments show
 that cells in a steady gradient can polarize in the direction of the
 gradient without extending pseudopods \cite{Parent:1999:CSD}. Thus
 cells must rely entirely on differences in the signal across the cell
 body for orientation.  Moreover, the timing between different
 components of the response is critical, because a cell must decide
 how to move before it begins to relay the signal. Analysis of a model
 for the cAMP relay pathway shows that a cell experiences a
 significant difference in the front-to-back ratio of cAMP when a
 neighboring cell begins to signal \cite{Dallon:1998:CAC}, which
 demonstrates that sufficient end-to-end differences for reliable
 orientation can be generated for typical extracellular signals.  An
 activator-inhibitor model for an amplification step in chemotactically
 sensitive cells was also postulated \cite{Meinhardt:1999:OCC}.  
 Amplification of
 small external differences involves a Turing instability in the
 activator-inhibitor system, coupled to a slower inactivator that
 suppresses the primary activation.  While this model reproduces some
 of the observed behavior, there is no biochemical basis for it; it is
 purely hypothetical and omits some of the major known processes.  A
 model that takes into account some of the known biochemical steps has
 been proposed more recently \cite{Krishnan:2005:MFD}.
 
 The objective of this paper is to derive equations for the
 population-level behavior of amoeboid cells such as Dd or leukocytes
 that incorporate details about the individual-based response to
 signals. We present several models with the increased complexity for
 which the derivation of population-level equations is possible.
 We show how experimentally-measured statistics can be obtained from the
 transport equation formalism. The paper is organized as follows.
 We discuss the classical chemotaxis
 description and summarize the state of the art of the derivation of
 macroscopic equations and population-level statistics from
 individual-based models in the remainder of this section. In Section
 \ref{secgensetup}, we establish the general setup for models of
 amoeboid cells and we present individual-based models which capture 
 the essential behavioral responses of eukaryotic cells. 
 In Section \ref{transport} we derive the
 macroscopic moment equations from the microscopic model and
 the dependence of the mean speed on the signal strength is studied. 
 Finally, we provide conclusions and the discussion of the presented 
 approaches in Section \ref{secdiscussion}.

\newpage

\subsection{Macroscopic descriptions of chemotaxis}

  The simplest description of cell movement in the presence of both
  diffusive and tactic components results by postulating that the flux
  of cells $\boldj$ is given by
\begin{equation}
\label{eq1.7a}
\boldj =    - D \nabla n  +  n  \boldu_c,
  \end{equation}
  where $n$ is the density of cells, $\boldu_c$ is the macroscopic
   chemotactic velocity and $D$ is the difusion constant.  The taxis
   is positive or negative according as $\boldu_c $ is parallel or
   anti-parallel to the direction of increase of the chemotactic
   substance $S$. Keller and Segel \cite{Keller:1970:ISM} postulated that the
   chemotactic velocity is given by $\boldu_c = \chi(S)\nabla S$ and
   then (\ref{eq1.7a}) can be written as
  \begin{eqnarray}
\label{eq1.7d}
 \boldj& =&  - D \nabla n  + n \chi(S) \nabla S
  \end{eqnarray}
  where $ \chi(S) $ is called the chemotactic sensitivity.  
  In the absence of cell division or death
  the resulting conservation equation for the cell density
  $n(\boldx,t)$ is
  \begin{equation}
\label{chemo}
\dfrac{\partial n}{\partial t} =  
\nabla \cdot( D \nabla n  -  n \chi(S) \nabla S)  
  \end{equation}
and this is  called a {\it classical chemotaxis equation}.
Unless the distribution of the chemotactic substance is fixed,
(\ref{chemo}) is coupled to an evolution equation for this substance,
and perhaps other governing variables.

Other phenomenological approaches to the derivation of the chemotactic
velocity have been taken.  For example, by approaching taxis from a
mechanical point of view, Pate and Othmer \cite{Pate:1986:DCS} derived 
the velocity
in terms of forces exerted by an amoeboid cell. Starting from Newton's
law, neglecting inertial effects, and assuming that the motive force
exerted by a cell is a function of the attractant concentration, they
showed how the chemotactic sensitivity is related to the rate of
change of the force with attractant concentration. In this formulation
the dependence of the flux on the gradient of the attractant arises
from the difference in the force exerted in different directions due
to different attractant concentrations.  Experimental support for this
comes from work of \cite{Varnum:1987:FOP}, who show that as many
pseudopods are produced down-gradient as up, but those up-gradient are
more successful in generating cell movement. We shall use a version of
the mechanical approach to taxis in a model described in the following
section.

The first derivation that directly relates the chemotactic velocity
to properties of individual cells is due to Patlak \cite{Patlak:1953:RWP},
who used kinetic theory arguments to express 
${\bf u}_c$ in terms of averages of the velocities and run times of
individual cells.  This approach was extended  by
Alt \cite{Alt:1980:BRW}, who showed that for a class of receptor-based
models the flux is approximately given by (\ref{eq1.7d}).
These approaches are based on velocity-jump processes, which lead to
transport equations of the form
\begin{equation}
\label{vjp}
\dfrac{\partial}{\partial t} p(\boldx,\boldv,t) + 
\boldv\cdot\nabla p(\boldx,\boldv,t) =
-\lambda p(\boldx,\boldv,t) + 
\lambda\int_V T(\boldv,\boldv') 
p(\boldx,\boldv'\!,t)\dboldv'. \;
\end{equation}
where $p(\boldx,\boldv,t)$ is the density of cells at 
position $\boldx\in\Omega\subset
\er^n$, moving with velocity $\boldv\in V\subset \er^n$ at time 
$t\geq 0$, $\lambda$ is the turning rate and kernel $T(\boldv,\boldv')$
gives the probability of a change in velocity from $\boldv^\prime$
to $\boldv$, given that a reorientation occurs \cite{Othmer:1988:MDB}.  
External signals enter either
through a direct effect on the turning rate $\lambda$ and the turning
kernel $T$, or indirectly via internal variables that reflect the
 external signal and in turn influence $\lambda$ and/or $T$. The
 first case arises when experimental results are used to directly  estimate
 parameters in the equation \cite{Ford:1992:SEQ}, but the latter
 approach is more fundamental. The reduction of (\ref{vjp}) to the
 macroscopic chemotaxis equations for the first case is done in
 \cite{Hillen:2000:DLT,Othmer:2002:DLT} and \cite{Chalub:2004:KMC}.

Some statistics of the density distribution in the first case, wherein
the external field modifies the turning kernel or turning rate
directly, can easily be derived and used to interpret experimental
data. To outline the procedure, we consider two-dimensional motion of amoeboid
cells in a constant chemotactic gradient directed along the positive
$x_{1}$ axis of the plane, i.e.
\begin{equation}
\nabla S =  \, \norm{\!\! \nabla S \!\!} \bolde_1,
\qquad
\mbox{where we denoted}
\qquad
\bolde_1 = [1,0].
\label{testgradient}
\end{equation}
Moreover, we assume that the gradient only influences
the turn angle distribution $T$; details of the
procedure are given in \cite{Othmer:1988:MDB}.  
We assume for simplicity that the individuals move with a constant
speed $s$. i.e. a velocity of an individual can be expressed as
$\boldv(\phi) \equiv s[\cos(\phi),\sin(\phi)]$ where $\phi \in [0,2 \pi).$
We assume that $T(\boldv,\boldv') \equiv T(\phi,\phi')$
is the sum of a symmetric probability distribution $h(\phi)$ and a
bias term $k(\phi)$ that results from the gradient of the chemotactic
substance.  Since the gradient is directed along the positive $x_{1}$
axis, we assume that the bias is symmetric about $\phi = 0$ and takes
its maximum there. Thus we write $T(\phi,\phi') = h(\phi -
\phi') + k(\phi)$ where $h$ and $k$ are normalized as follows.
\begin{equation}
\int_{0}^{2\pi} h(\phi) \dphi  =  1 \qquad \qquad \int_{0}^{2\pi}
k(\phi) \dphi  =  0 
\label{72}
\end{equation}
Let $p(\boldx,\phi,t)$
be the density of cells at position $\boldx \in \er^2$, moving 
with velocity $\boldv(\phi) \equiv s[\cos(\phi),\sin(\phi)]$, 
$\phi \in [0,2 \pi)$, at time 
$t\geq 0$. The statistics of interest are the mean location of cells
${\boldX}(t)$, their mean squared
displacement   ${\cal D}^{2}(t)$, and their mean velocity $ {\boldV}(t)$,
which are defined as follows.
$$
    {\boldX}(t) = \frac{1}{N_{0}} \int_{\er^2}
	\int_{0}^{2\pi}
	\boldx p(\boldx,\phi,t)\,\,\dphi \,\dboldx,
$$
$$
    {\cal D}^{2}(t) = \frac{1}{N_{0}} 
	 \int_{\er^2}\int_{0}^{2\pi}
         \parallel\!\!\boldx\!\!\parallel^2 
	 p(\boldx,\phi,t)\,\,\dphi \,\dboldx, 
$$
$$
    {\boldV}(t) = \frac{1}{N_{0}}
     \int_{\er^2}\int_{0}^{2\pi} 
     \boldv(\phi)  p(\boldx,\phi,t)\,\, \dphi \,\dboldx, 
$$
$$
    {\cal B}(t)
    =\frac{1}{N_{0}} \int_{\er^2}\int_{0}^{2\pi}(\boldx \cdot \boldv(\phi))
    p(\boldx,\phi,t)\,\, \dphi \,\dboldx, 
$$      
where $N_0$ is the total number of individuals present and ${\calB}(t)$ 
is an auxiliary variable that is needed in the analysis. Two
further quantities that arise naturally are the taxis coefficient
$\chi $, which is analogous to the chemotactic sensitivity defined
earlier because it  measures the response to a directional signal, and
the persistence index $\psi_d$. These are defined as
\begin{equation}
\chi \equiv \int_{0}^{2\pi} k(\phi) \cos \phi \, \dphi \qquad
      \mbox{and } \qquad \psi_{d} = 
2\int_{0}^{\pi}h(\phi)\cos\phi \, \dphi. 
\label{43}
\end{equation}
The persistence index  measures the tendency of a cell to continue in the current
direction.  Since we have assumed that the speed is constant, we must
also assume that $\chi$ and $\psi_{d}$ satisfy the
relation $\chi < 1 - \psi_{d}$, for otherwise the former assumption is
violated (\cf (\ref{xvsol})).

  One can now show, by taking moments of (\ref{vjp}), using
(\ref{72}) and symmetries of $h$ and $k$, that the moments satisfy 
the following evolution equations
\cite{Othmer:1988:MDB}. 
\begin{eqnarray}
 \label{stat1}
   \dfrac{\dd {\boldX}}{\dt} = {\boldV}
    \qquad
    &
    \qquad
    \dfrac{\dd {\boldV}}{\dt} = - \lambda_{0} {\boldV} + 
    \lambda \chi s \bolde_1 \\
    \dfrac{\dd{\calD}^{2}}{\dt} = 2 {\calB}
    \qquad
   &
    \qquad
    \dfrac{\dd {\calB}}{\dt} = s^2 - \lambda_{0} {\calB} + 
    \lambda \chi s X_{1}
\label{stat2}
\end{eqnarray} 
where $\lambda_{0} \equiv \lambda(1 - \psi_{d})$.
The solution of (\ref{stat1}) subject to zero initial data is 
\begin{equation}
    {\boldX}(t)
         =s C_I
          \left(t - \dfrac{1}{\lambda_{0}}(1 - e^{-\lambda_{0}t})\right)
	  \bolde_1,
\qquad {\boldV}(t)  = s C_I (1 - e^{-\lambda_{0}t}) \, \bolde_1
\label{xvsol}
\end{equation}
where  $C_I \equiv \chi/(1 - \psi_{d})$ 
is sometimes called the chemotropism index.
Thus the mean velocity of cell movement is parallel to the direction of the
chemotactic gradient and approaches 
${\boldV}_{\infty} = s \, C_I \bolde_1$
as $t \rightarrow \infty$. Thus the asymptotic mean  speed is the cell
speed  decreased by the factor $C_I$.

A measure of the fluctuations of the cell path around the expected value
is provided by the  mean square deviation, which is defined as
\begin{equation}
\sigma^{2}(t)
=
\frac{1}{N_0}
\int_{\er^2} \int_{0}^{2\pi}
\parallel \!\!\boldx - {\boldX}(t)\!\!\parallel^{2} 
p(\boldx,\phi,t)\, \dphi \dboldx
= {\calD}^{2}(t)  -  \norm{\!\!{\boldX}(t)\!\!}^2.
\label{79}
\end{equation}
Using (\ref{stat1}) -- (\ref{stat2}), one also finds a differential equation
for $\sigma^{2}$. Solving this equation, we find
\[
\sigma^{2} \sim
\displaystyle{\frac{2s^{2}}{\lambda_{0}}}
\left\{(1 - C_I^2)t
+  
\frac{1}{\lambda_{0}}  
\left(
\frac{5}{2}
C_I^2 - 1
\right)
\right\}
\qquad
\mbox{as}
\qquad
t \to \infty
\]
and from this one can extract the diffusion coefficient as 
$$ 
D = \dfrac{2s^{2}}{\lambda_{0}}(1 - C_I^2).  
$$ 
Therefore if the effect of an external gradient can be quantified
experimentally and represented as the distribution $k$, the
macroscopic diffusion coefficient, the persistence index, and the
chemotactic sensitivity can be computed from measurements of the mean 
displacement, the asymptotic speed and the mean-squared displacement.

However, it is not as straightforward to derive directly the
macroscopic evolution equations based on detailed models of signal
transduction and response. Suppose that the internal dynamics that
describe signal detection, transduction, processing and response are
described by the system
\begin{equation}
\dfrac{\dd \boldy}{\dt} = \boldf(\boldy, S) 
\label{intradyn}
\end{equation}
where $\boldy \in \er^{m}$ is the vector of internal variables and $S$
is the chemotactic substance ($S$ is extracellular cAMP for Dd
aggregation).  Models that describe the cAMP transduction pathway
exist \cite{Martiel:1987:MBR,Tang:1994:GPB,Tang:1995:EOW}, but for
describing chemotaxis one would have to formulate a more detailed
model.  The form of this system can be very general but it should
always have the ``adaptive'' property that the steady-state value 
(corresponding to the constant stimulus)
of the appropriate internal variable (the ``response regulator'') is
independent of the absolute value of the stimulus, and that the steady 
state is globally attracting with respect to the positive cone of $\er^{m}$.

We showed earlier that for non-interacting walkers the internal
dynamics can be incorporated in the transport equation as follows
\cite{Erban:2004:ICB}. Let $p(\boldx,\boldv,\boldy,t)$ be the density 
of individuals
in a $(2N+m)-$dimensional phase space with coordinates $[\boldx,\boldv,\boldy]$,
where $\boldx
\in \er^N$ is the position of a cell, $\boldv \in V \subset \er^N$ is its
velocity and $\boldy \in Y
\subset \er^m$ is its internal state, which evolves according to
(\ref{intradyn}).  The evolution of $p$ is governed by the transport
equation 
\begin{equation}
 \frac{\partial p}{\partial t} + \nabla_\boldx \cdot \boldv p + 
 \nabla_\boldy \cdot \boldf
p = -\lambda(\boldy) p + \int_{V} \lambda(\boldy) 
T(\boldv,\boldv^\prime\!,\boldy)
p(\boldx,\boldv^\prime,\boldy,t) \dboldv^\prime
\label{vjpint}
\end{equation}
where, as before, we assume that the random velocity changes are the
result of a Poisson process of intensity $\lambda(y)$.  The kernel
$T(\boldv,\boldv^\prime\!,y)$ gives the probability of a change in velocity 
from $\boldv^\prime$ to $\boldv,$ given that a reorientation occurs.  
The kernel $T$
is non-negative and satisfies the normalization condition $\int_V
T(\boldv,\boldv^\prime,y) \dboldv = 1.$ To connect this with the chemotaxis 
equation
(\ref{chemo}), we have to derive an evolution equation for the
macroscopic density of individuals
\begin{equation}
n(\boldx,t) = \int_Y \int_V p(\boldx,\boldv,\boldy,t) \dboldv\dboldy.
\label{rom202a}
\end{equation}
The problem turns out to be tractable for systems that execute
``run-and-tumble'' motion, such as {\em E. coli}.  To illustrate this,
assume for simplicity that the motion is restricted to 1D, the signal
is time-independent, the speed $s$ is constant, and the turning phase is
neglected; the general cases are treated elsewhere
\cite{Erban:2004:ICB,Erban:2005:STS}.  Let $p^+$ (resp. $p^{-}$) be
the density of individuals moving to the right (resp. left). Then
(\ref{vjpint}) leads to a telegraph process described by the
hyperbolic system
\begin{equation}
\frac{\partial p^+}{\partial t}
+
s\frac{\partial p^+}{\partial x}
+
\sum_{i=1}^m
\frac{\partial}{\partial y_i}
\left[ f_i(\boldy,S)
p^+ \right]
=
\lambda (\boldy)
 \left[ - p^+ + p^- \right],
\label{rom301}
\end{equation}
\begin{equation}
\frac{\partial p^-}{\partial t}
-
s
\frac{\partial p^-}{\partial x}
+
\sum_{i=1}^m
\frac{\partial}{\partial y_i}
\left[ f_i(\boldy,S)
p^- \right]
=
\lambda (\boldy)
 \left[ p^+ -  p^- \right].
\label{rom302}
\end{equation}
The essential components of the internal dynamics in the bacterial
context are fast excitation, followed by slower adaptation and return
to the basal turning rate, and these aspects are captured in the
system \cite{Othmer:1998:OCS}
\begin{equation}
\frac{dy_1}{dt} = \frac{g(S(x)) - (y_1 + y_2)}{\tau_e} \qquad
\mbox{and} \qquad
\frac{dy_2}{dt} = \frac{g(S(x)) - y_2}{\tau_a}.
\label{rom81}
\end{equation}
Here $g$ encodes the first step of signal
transduction, $S$ is the chemoattractant, and $\tau_e$ and $\tau_a$
are time constants for excitation and adaptation, respectively. The
component $y_1$ adapts perfectly to constant stimuli, {\em i.e.}, the
steady state response is independent of the magnitude of the stimulus
$S$.  To obtain a macroscopic limit equation for the total density
$n(x,t)$ we incorporate the variables $y_i$ into the state and derive
a system of four moment equations for various densities and fluxes
\cite{Erban:2004:ICB}. 
Assuming that the turning rate has the form $\lambda(y)=\lambda_0-b y_1$,
for $\lambda_0 > 0$, $b>0$,
we show that this system reduces to the
classical chemotaxis equation for large times
\begin{equation}
\frac{\partial n}{\partial t}
=
\frac{\partial}{\partial x}
\left(
\frac{s^2}{2\lambda_0}
\frac{\partial n}{\partial x}
-  
\left[
\frac{b s^2 \tau_a g^\prime(S(x))} {\lambda_0 (1 + 2 \lambda_0
\tau_a)(1 + 2 \lambda_0 \tau_e)}\right] S^\prime(x) n\right)
\label{rom238}
\end{equation}
\noindent 
where
the chemotactic sensitivity is given explicitly in terms of parameters
that characterize signal transduction and response.  We have only used
the simplified dynamics (\ref{rom81}) to obtain the macroscopic
chemotactic sensitivity, but this model  captures the
essential aspects for bacterial taxis
\cite{Spiro:1997:MEA,Erban:2004:ICB}. An open problem is
how one extracts the elementary processes of excitation and adaptation
from a complex network of the type used for signal transduction in
{\em E. coli}. Finally, let us note that the global existence results for 
(\ref{vjpint}) which is coupled with the evolution equation for the 
extracellular signal were recently given in \cite{Erban:2006:GER}.

Equation (\ref{rom238}) was derived for cells such as bacteria, that
use the ``run-and-tumble'' strategy, and our objective in this paper
is to attempt a similar reduction of the transport equation to a
chemotaxis equation for more complex amoeboid eukaryotic cells.  In
the following section we introduce the general setup for studying
amoeboid taxis. Then we study several ``caricature'' or ``cartoon''
models for amoeboid chemotaxis with the objective of deriving
macroscopic population-level equations in each case.  We start with a
model which can capture interesting features of eukaryotic motility
without introducing additional internal state variables, and then add
internal state variables to the model.

\section{Amoeboid taxis with internal variables} 

\label{secgensetup}

A fundamental assumption in the use of velocity-jump processes 
\cite{Othmer:1988:MDB} to
describe cell motion is that the jumps are instantaneous, and
therefore the forces are Dirac distributions. This approximates the case
in which very large forces act over very short time intervals, and
even if one incorporates a resting or tumbling phase, as was done in
\cite{Othmer:2002:DLT}, the macroscopic description of motion is unchanged. This is
appropriate for the analysis of bacterial motion (and other systems
that use a ``run-and-tumble'' strategy), as summarized above, since
the effect of the external signal is to change the rotational behavior
of the flagella, and not, so far as it is understood, to affect the
force generation mechanism itself. However, the situation is very
different when analyzing the movement of crawling cells, for here the
control of the force-generation machinery is an essential component of
the response. While amoeboid cells such as Dd extend pseudopods
``randomly'' in the absence of signals, the direction of extension is
tightly controlled in the presence of a directed external signal, and
the direction in which forces are exerted on the substrate is 
controlled via the location of contacts with the substrate. Therefore
it is appropriate to incorporate the force-generation machinery as
part of the internal state, and as a first step we condense this all
into a description of how the force exerted by a cell on its
surroundings (and {\em vice-versa}) depends on the external signal. In
reality amoeboid cells are also highly deformable, and a complete
theoretical treatment of taxis at the single cell level has to take
this into account. This is currently under investigation but will not
be pursued here; instead we only describe the motion of the centroid
of the cell. However, the following framework is sufficiently general
to allow distributed internal variables within a cell. 

Hereafter we use $\boldy$ as it appears in (\ref{intradyn1}) to denote the
chemical variables involved in signal transduction, control of actin
polymerization, etc, and we denote the force per unit mass on the  centroid
of a cell by ${\cal F}(\boldx,\boldv,\boldy)$. Therefore the internal state 
equations are given by
\begin{equation}
\dfrac{\dd\boldy}{\dt} = {\cal G}(\boldy, S) 
\label{intradyn1}
\end{equation}
and the velocity evolves according to 
\begin{equation}
\dfrac{\dboldv}{\dt} = {\cal F}(\boldx,\boldv,\boldy). 
\label{intradyn2}
\end{equation}
Here ${\cal G}: \mathbb Y \times \mathbb S \rightarrow \mathbb Y$ is in general
a mapping between suitable Banach spaces and 
${\cal F}: \mathbb R^N
\times \mathbb R^N \times \mathbb Y \to \mathbb R^N$
where $N=1,2,$ or $3$ is the dimension of the physical space. 
This generality is needed
because the variable $\boldy$ can include quantities that depend on the
location in the cell or on the membrane, and  which  may, for example,
satisfy a reaction-diffusion equation or another evolution equation.

The cell is therefore described by the position and velocity of its
centroid, and the internal state $\boldy \in \mathbb Y.$ In some important
cases described later there is a projection ${\cal P}: \mathbb Y \to \mathbb
Z \subset \mathbb Y$ from $\mathbb Y$ onto a suitable finite-dimensional
subspace $\mathbb Z$, obtained for example by considering the first few
modes in a suitable basis for $\mathbb Y$, such that
\begin{equation}
{\cal P} ({\cal G}(\boldy,S)) =  \boldG(\boldz,S) 
\quad \mbox{and} \quad
{\cal F} (\boldx,\boldv,\boldy) =  \boldF(\boldx,\boldv,\boldz),\quad 
\mbox{~ where~} \quad \boldz \equiv {\cal P} \boldy. 
 \label{requirements}
\end{equation}
Here $\boldG(\cdot,S): \mathbb Z \to \mathbb Z$ and 
$\boldF(\cdot,\cdot,\cdot): 
\er^N \times \er^N \times \mathbb Z \to \er^N$ are mappings
between finite-dimensional spaces. The first equality defines the
function $\boldG$, though it may of course be difficult to find when
${\cal G}$ is nonlinear. The function $\boldF$ is explicitly given by
the second equality when the reduction is possible.

Given a suitable choice of the
projection ${\cal P}$, we can reduce the infinite-di\-men\-sio\-nal system
(\ref{intradyn1}) -- (\ref{intradyn2}) to the following set of
ordinary differential equations in finite dimensions for the description
of individual cells. 
\begin{eqnarray}
\label{zequation}
\dfrac{\dd\boldz}{dt} &=& \boldG(\boldz,S) \\
\dfrac{\dboldv}{\dt} &=& \boldF(\boldx,\boldv,\boldz) 
\label{vequation}
\end{eqnarray}
 Next, let $p(\boldx,\boldv,\boldz,t)$ be the
density  of individuals which are at point
$\boldx$, with velocity $\boldv$ and with the vector of reduced
internal variables $\boldz$;  then the transport equation 
(\ref{vjpint}) can be written in the form 
\begin{equation}
 \frac{\partial p}{\partial t} + \nabla_\boldx \cdot \boldv p + \nabla_v \cdot
\boldF p  + \nabla_z \cdot \boldG 
p = -\lambda(\boldz) p + \int_{V} \lambda(\boldz) T(\boldv,\boldv'\!,\boldz)
p(\boldx,\boldv^\prime\!,\boldz,t) \dboldv^\prime.
\label{vjpint1}
\end{equation}
A crucial assumption for using the transport equation
formalism is that the projection ${\cal P}$ exists; at present we
do not know how to extend this framework to an infinite-dimensional
manifold.  Examples of models for which the projection ${\cal P}$ can
be found will be given in the following sections, and in these cases  we
can  use  (\ref{vjpint1}) as the starting point for
obtaining macroscopic equations. As described earlier, the right-hand
side models the instantaneous changes of direction of motion, and in
the present context we use this to describe the small fluctuations
due to random ``errors" in the sensing of the signal and possibly to
an intrinsic mechanism for  random exploration of the local
environment. Tranquillo and Lauffenburger \cite{Tranquillo:1987:SML} 
developed a model of amoeboid
movement that focuses specifically on the stochastic component. 

A natural question is what can be done if a suitable projection ${\cal
P}$ is not easily computed, or if the explicit form of $\boldG$ is
impossible to obtain because of the complexity of the mapping $\cal
G$.  In some cases it may still be possible to describe the
macroscopic-level dynamics by the evolution of a few slow variables,
and by using computational equation-free methods which are currently
being developed, to obtain populational level quantities without
explicitly deriving the macroscopic equations (see
\cite{Kevrekidis:2003:EFM,Erban:2006:EFC,Erban:2006:GRN}
and references there), using either the full model of the amoeboid cell
or the best available reduction of it.

In the remainder of the paper we give examples of the reduction of
 (\ref{intradyn1}) -- (\ref{intradyn2}) to the form (\ref{zequation})
 -- (\ref{vequation}) and the derivation of macroscopic equations via
 the transport equation (\ref{vjpint1}), in order to understand how
 the population-level dynamics depends on the characteristics of the
 individual behavior.  We start with a motivating example in which we
 further reduce the system (\ref{zequation}) -- (\ref{vequation}) by
 assuming that $\dim \mathbb{Z} =1$ and that the function
$\boldG$ transduces the signal directly, {\em i.e.,} $\boldz \equiv S$.

\subsection{A motivating example} 
\label{secdirection}

\begin{figure}[hbt!]
\centerline{ 
\psfig{file=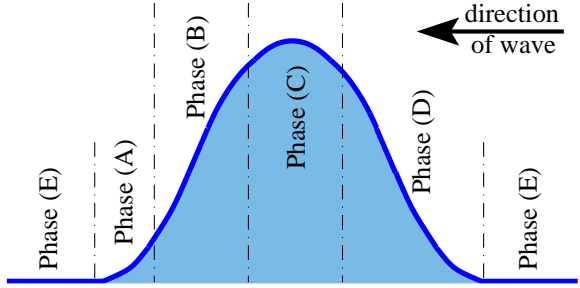,height=1.5in} 
} \caption{{\it The notation for the different phases of the wave of
  chemoattractant seen by a cell at a fixed  spatial position, as a
  function of time. The horizontal axis is the time and the vertical
  axis is the amplitude of the signal.}}
\label{figwave}
\end{figure}

To illustrate how the effect of acceleration of the cell can enter
into the macroscopic equations, we consider the example of motion of a
cell in the plane in response to a wave of a chemotactic substance.
The typical response of Dd or leukocytes to a pulse-like wave of the
chemoattractant can be divided into several phases depending on the
position of the cell relative to the wave \cite{Geiger:2003:HPL}. In
Figure \ref{figwave} we distinguish five different phases - denoted
(A) -- (E).
Before the wave arrives at the cell, there are no directional cues in
the environment and the cell extends pseudopods in all
directions -- Phase (E). When the wave arrives the cell experiences an 
increasing temporal gradient at all points of its surface and can detect 
a front-to-back spatial gradient over its length (where front denotes
the direction of the oncoming wave), which causes it to polarize in 
the direction of the oncoming wave.  This is  Phase (A) in Figure
\ref{figwave}. In Phase (B) lateral pseudopod formation is
suppressed and the cell moves more-or-less directly towards the
aggregation center at a speed of 10-20 $\mu$m/min.  In natural cAMP
waves the cAMP concentration at the peak of the wave is high enough
that in Phase (C) the cells stop translocating and depolarize. In Phase
(D) the temporal gradient is negative, although the spatial gradient is
positive in the outgoing direction, and the cell begins to form
pseudopods in all directions. This is presumably due to slow adaptation to
the decreasing cAMP signal, and as we shall see, if it is too fast the
cells may reverse direction and follow the outgoing wave. In Phase
(E), there is no extracellular signal present and there is not net
movement of cells. This last phase is not described in
\cite{Geiger:2003:HPL} but it is of interest to include this to
describe the motion in the absence of a stimulus. Formal rules used in
the context of an individual-based model of Dd aggregation show that
population-level aspects of chemotaxis such as stream formation can be
reproduced if the foregoing phases are properly incorporated
\cite{Dallon:1997:DCM}. How to incorporate these characteristics into
a continuum description is the question addressed here.

The following example is not meant to provide a realistic description
of taxis, but rather to motivate the analysis done later. In a coarse
or high-level description of movement in response to signals,
information carried by the external signal detected by a cell is
transduced through the intracellular signaling network, and during
deterministic turns the velocity of the cell follows the external
gradient with some delay. We write Newton's law for the motion of the
centroid as
\begin{equation}
\dfrac{\dboldv}{\dt} = \dfrac{\boldg(\nabla S) - \boldv}{\tau_v(S)}
\label{velin1}
\end{equation}
where we assume that the relaxation (or adaptation) time $\tau_v(S)$ 
is a functional of $S.$ Term $\boldg(\nabla S)$ can be interpreted 
in terms of the force generated from the extracellular signal.
Typically $\boldg(\nabla S)$ vanishes at zero, is monotone
increasing, and saturates for large $\nabla S$. 
The dependence of $\tau_v(S)$ on $S$ could arise, for instance, 
from different responses of the intracellular dynamics to 
increasing and decreasing signals; or
from alterations in the adhesion sites between cell and  substrate. 
In earlier work the turning behavior was  incorporated via rules
\cite{Dallon:1997:DCM}, rather than via an equation of motion 
such as (\ref{velin1}). 

To demonstrate that this model can capture some of the salient
features of Dd aggregation in response to cAMP waves from a pacemaker
center, we present the results of cell-based numerical simulations
that use (\ref{velin1}) for the velocity, given a suitable choice of
$\tau_v(S)$.  We consider a two-dimensional disk (corresponding to a Petri
dish) of radius $5$ mm, and we specify a periodic source of cAMP waves
at the center of the domain.  The period of the waves is seven
minutes, their speed is $400\mu{}\mbox{m/min}$, and the maximal speed
of a cell is about 20 $\mu$m per minute, all of which are chosen to
approximate natural waves in a Dd aggregation field. More precisely,
we choose 
$\boldg(\nabla S) = s_0 \nabla S/(c_s + \norm{\!\!\nabla S\!\!})$
where $s_0 = [20 \mu\mbox{m/min}]$, and $c_s$ measures the sensitivity
of the signal transduction mechanism. In the numerical examples we
choose a wave with maximum $\norm{\!\!\!\nabla S\!\!\!}$ equal to $1$
mm$^{-1}$ and $c_s = 10^{-4} \mbox{mm}^{-1}$. Initially the cells are
distributed uniformly and we investigate under what conditions the
cells aggregate at the source of the waves of chemoattractant $S$.  We
consider the following two choices for the dependence of $\tau_v(S)$ on the
external field.

\begin{figure}[hbt!]
\centerline{
\psfig{file=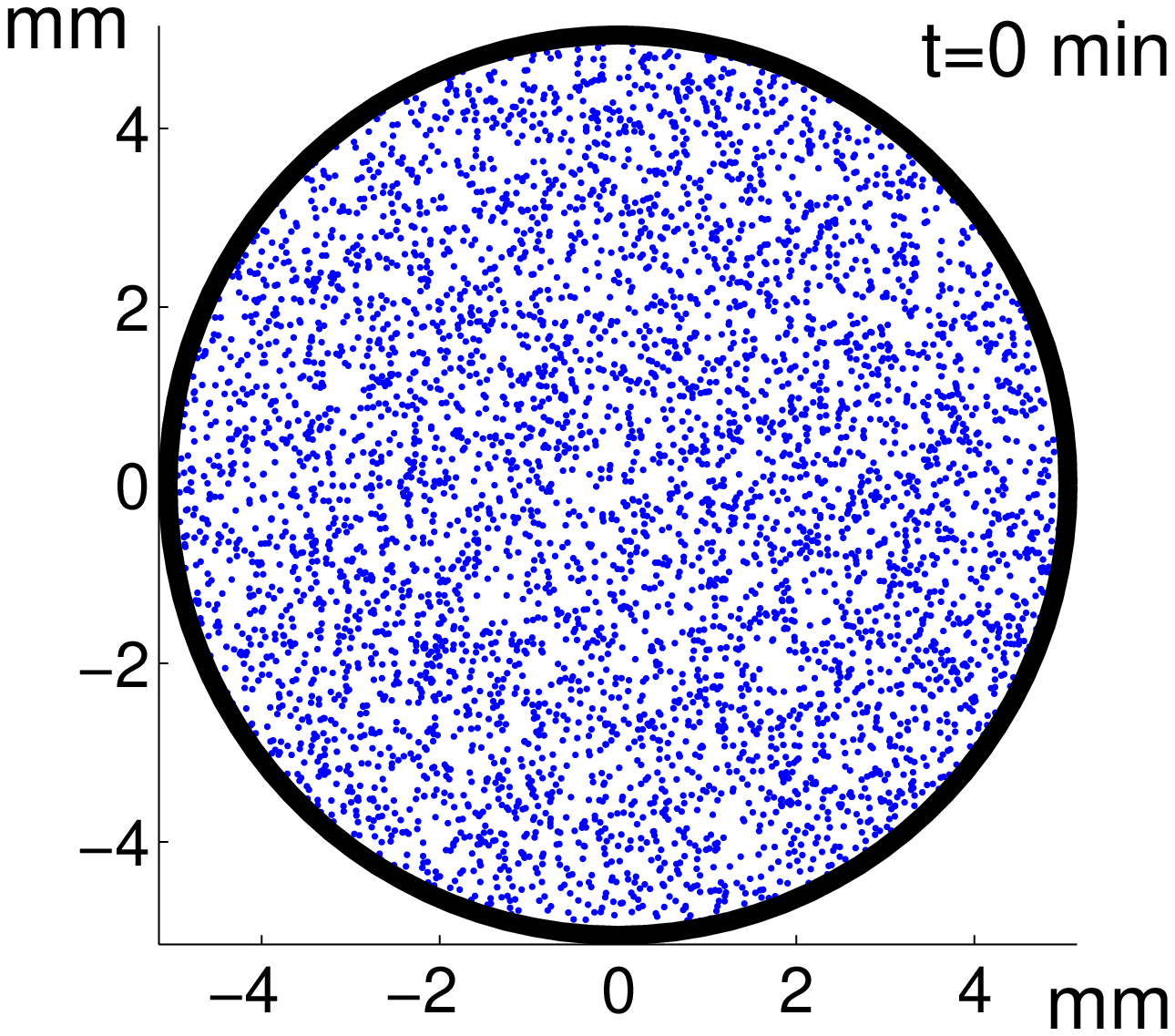,width=2.25in}
\psfig{file=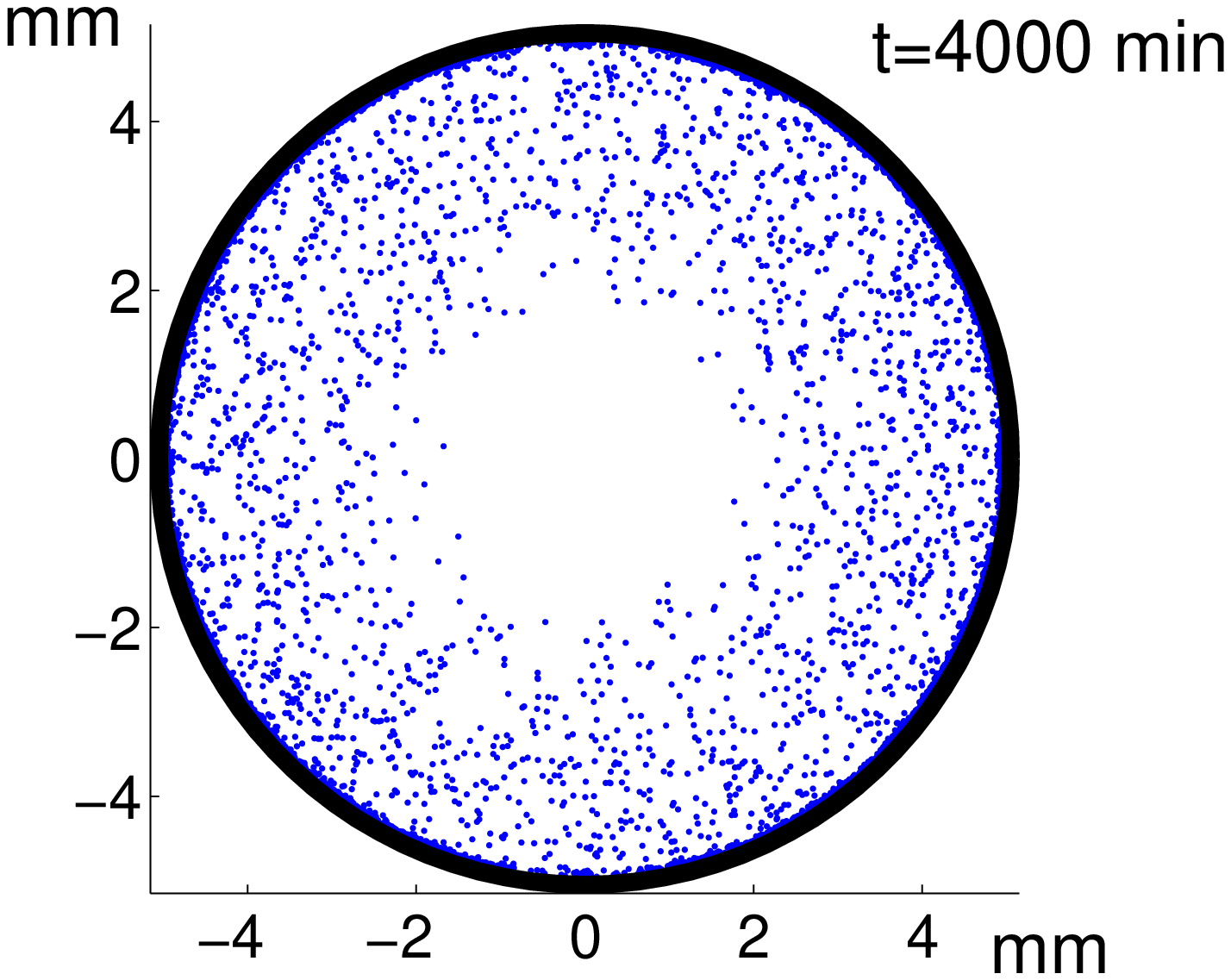,width=2.25in}
}
\caption{{\it Simulation of \/ $5000$ cells that move according to  
$(\ref{velin1})$
when the relaxation time $\tau_v(S)$ is constant. We plot the positions 
of cells at $t=0$ (left) and at $t=4000$ min (right).}}
\label{mod1figB}
\end{figure}

\smallskip\noindent
{\bf (1)} $\tau_v(S)$ is a constant independent of the signal (\cf
Figure \ref{mod1figB}). In this case there is no aggregation, and in
fact, cells move to the boundary of the Petri dish. This is not
surprising, because cells move in the direction of the increasing
gradient of the attractant, and cells first move toward the source and
then turn around. Although the wave is symmetric, the cell movement
creates a cellular ``Doppler effect" in that there is an asymmetry in
the time the cell detects the inward-directed gradient at the front of
the wave versus the time it sees the receding wave. Thus in every
cycle it moves away from the source longer than it moves toward it,
and cells eventually accumulate at the boundary.

\begin{figure}[hbt!]
\centerline{
\psfig{file=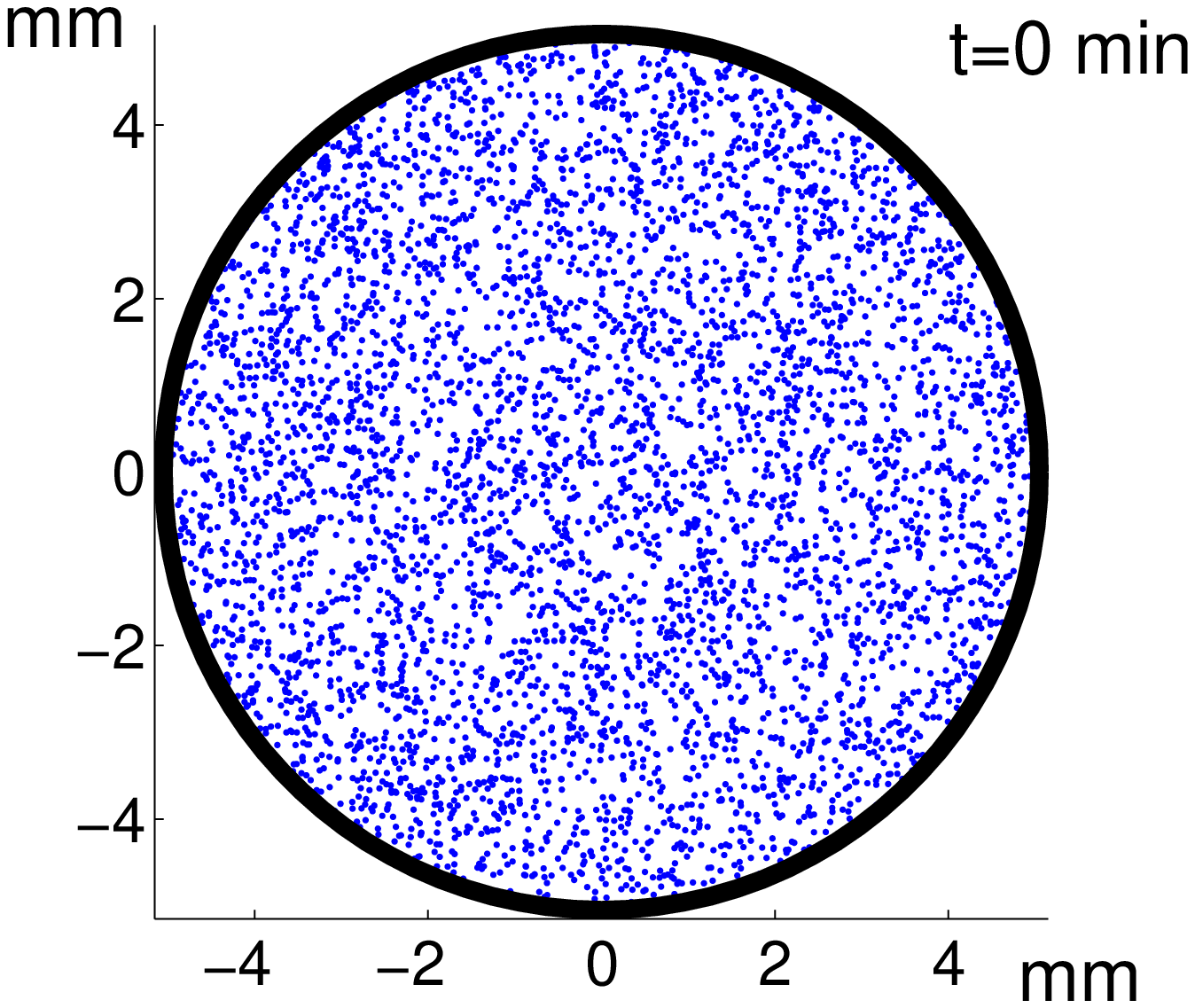,width=2.25in}
\psfig{file=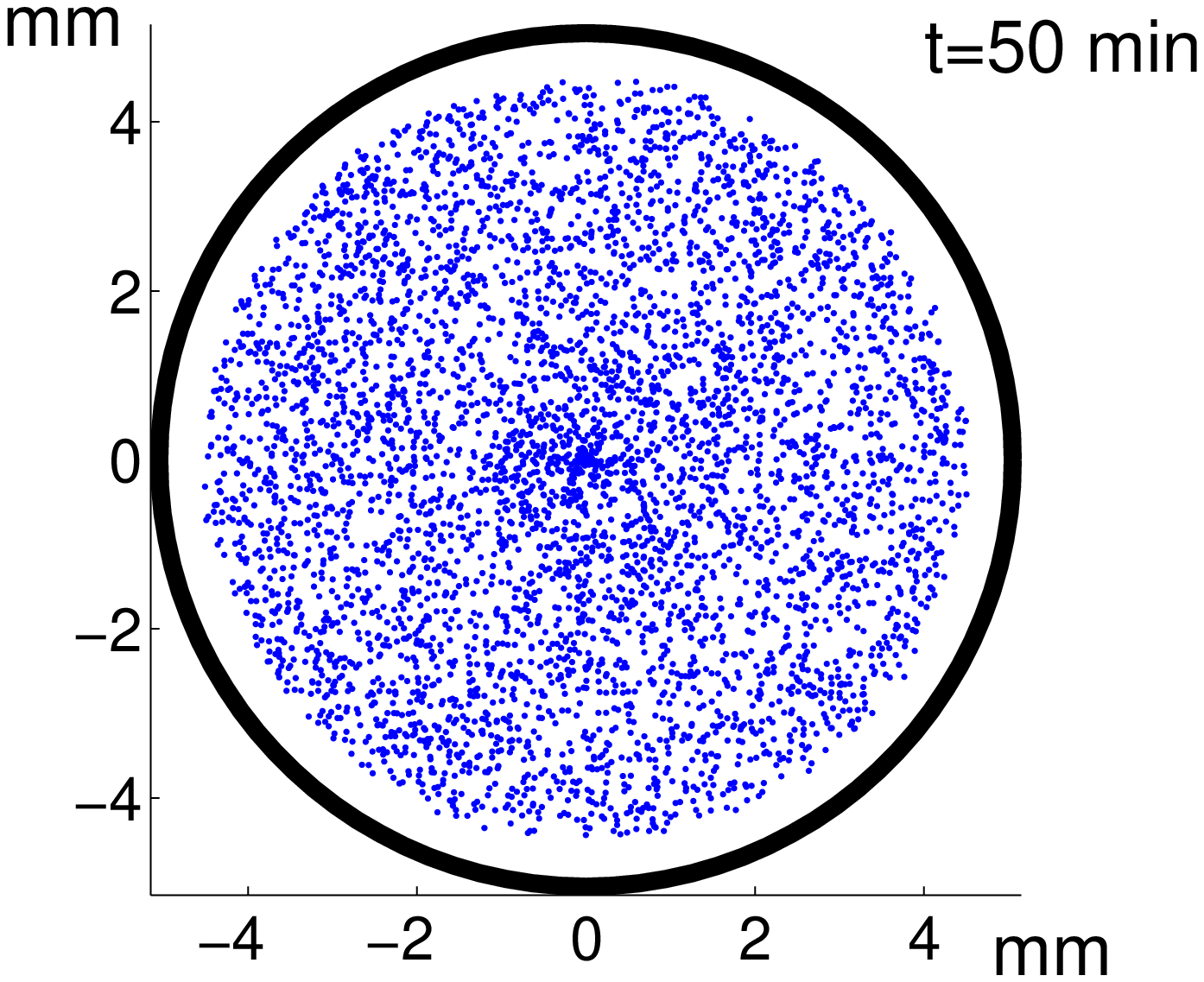,width=2.25in}
}
\centerline{
\psfig{file=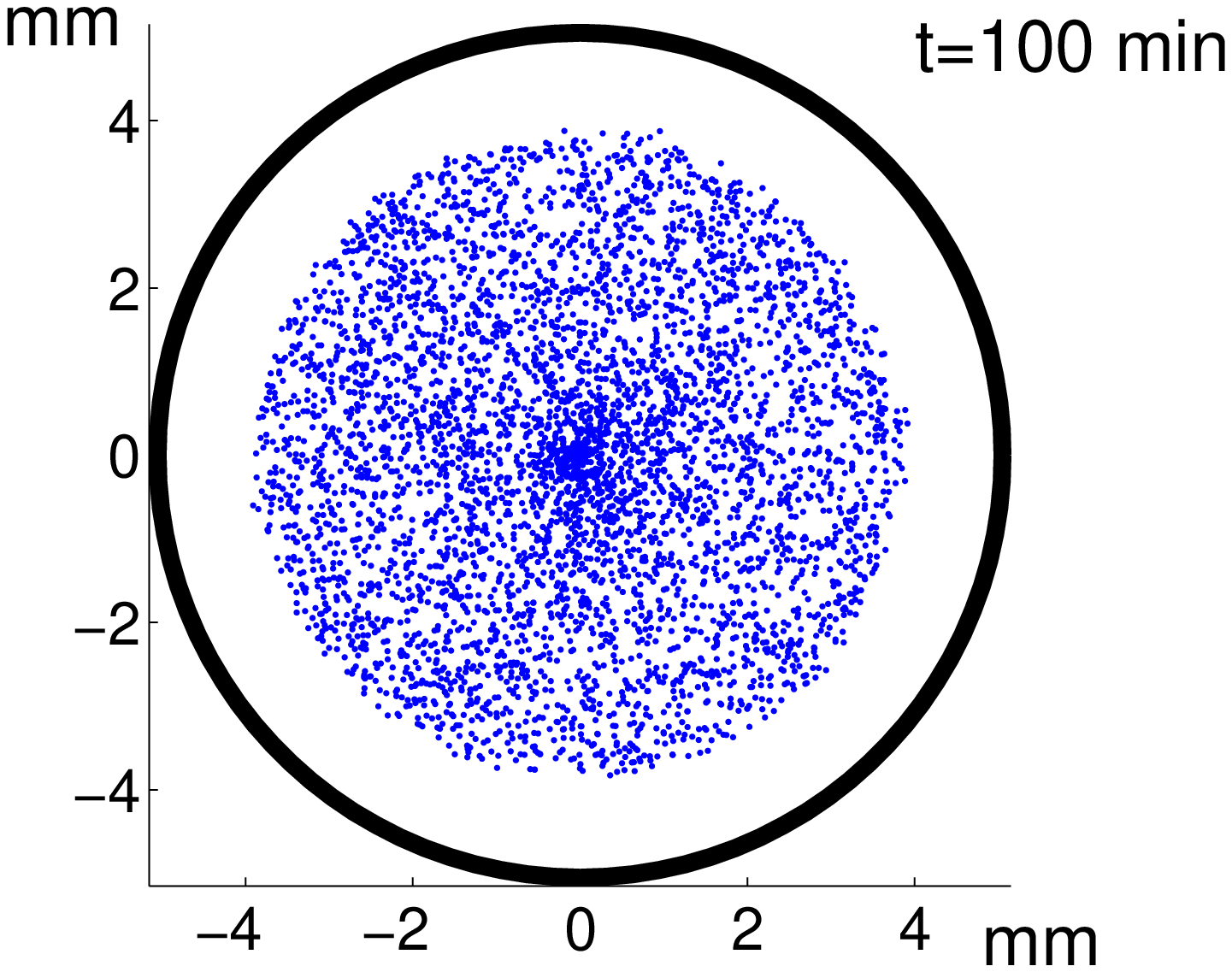,width=2.25in}
\psfig{file=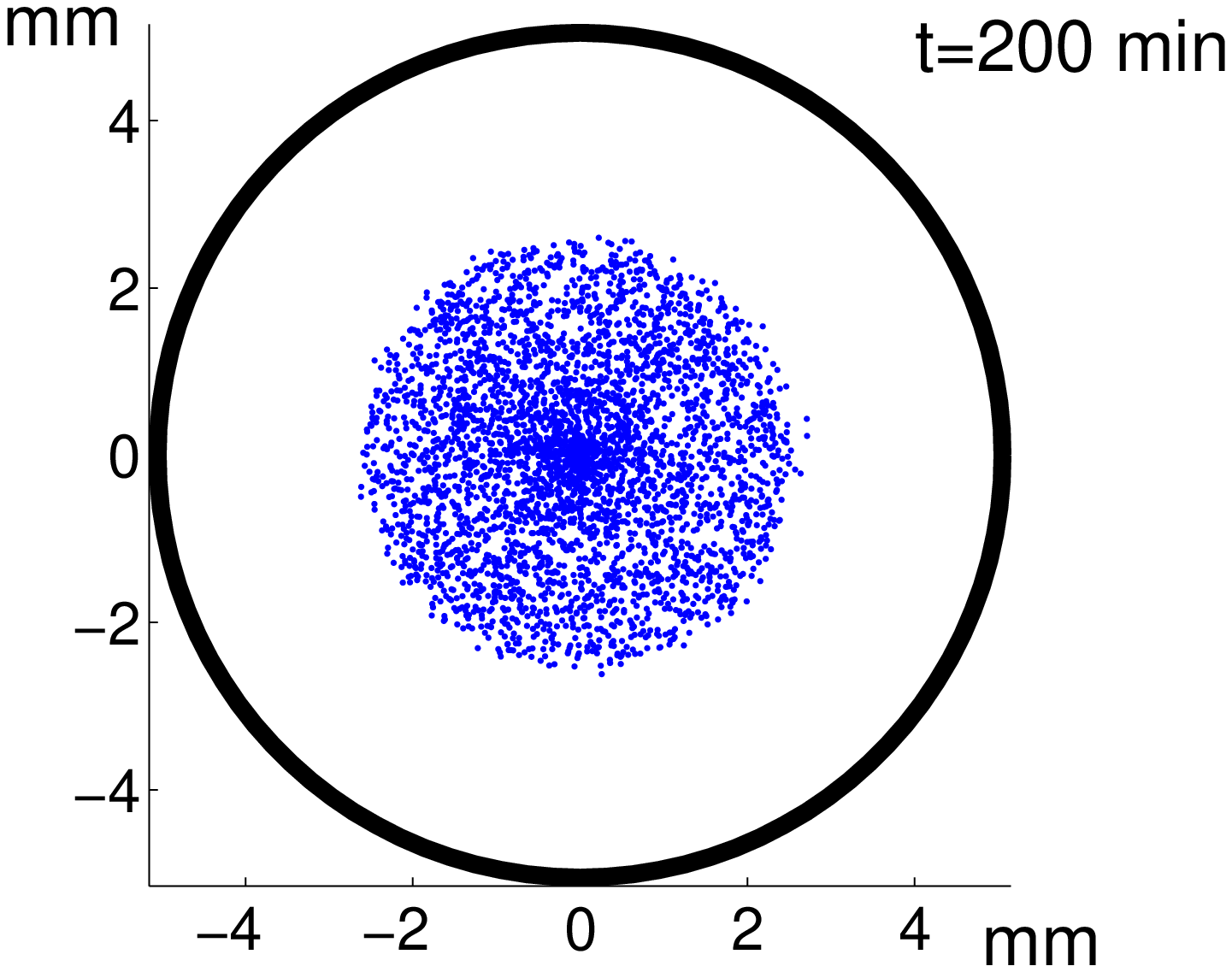,width=2.25in}
}
\centerline{
\psfig{file=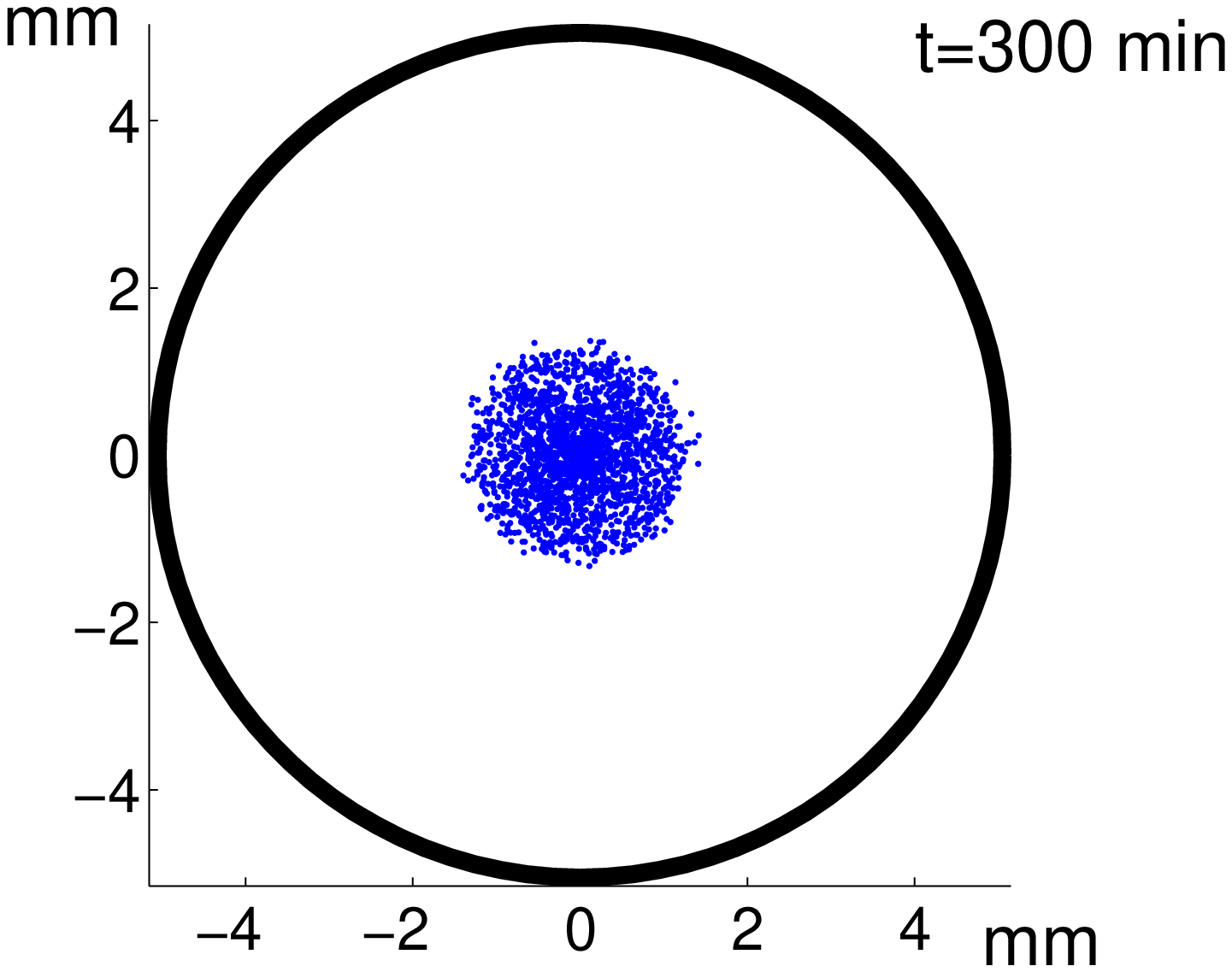,width=2.25in}
\psfig{file=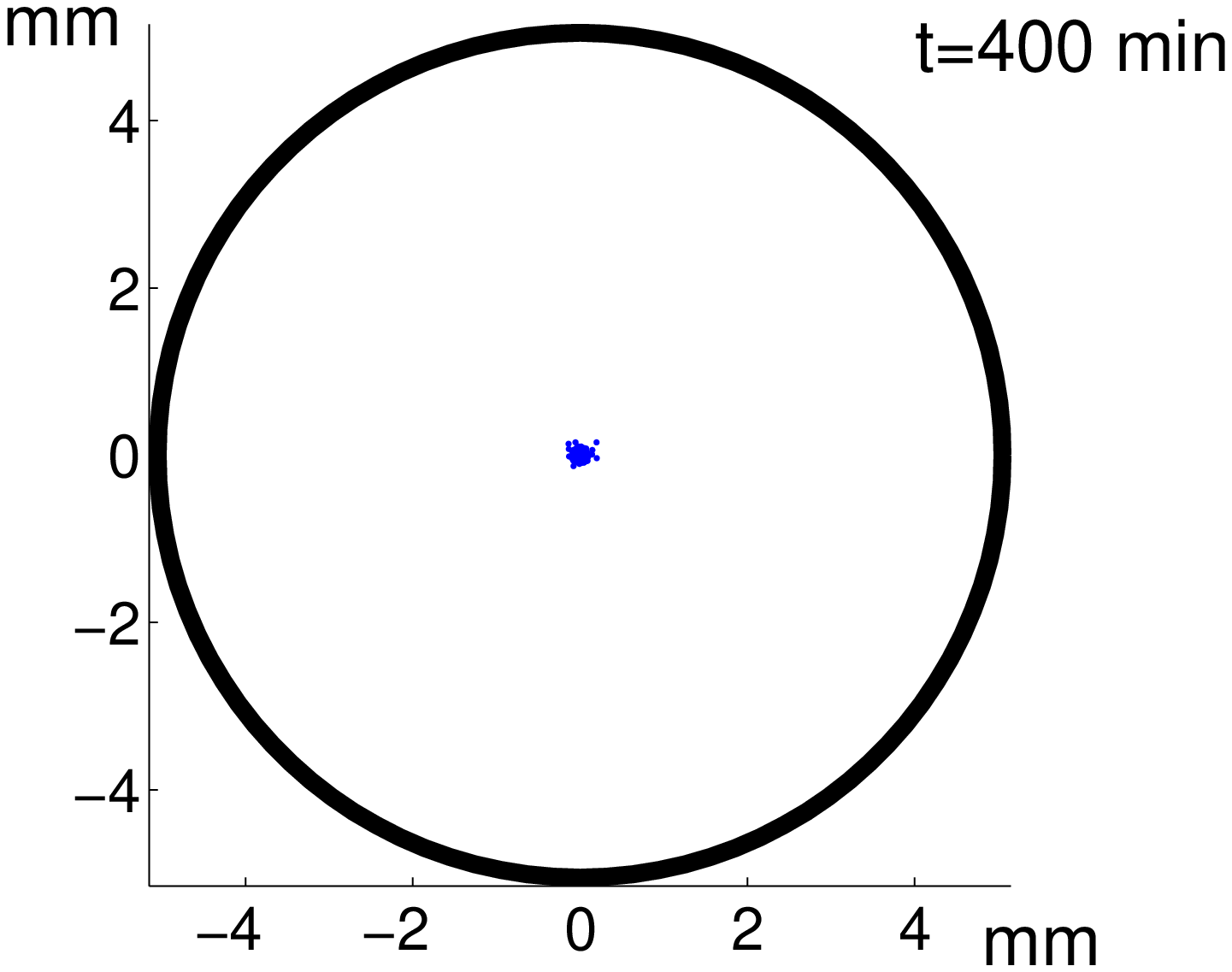,width=2.25in}
}
\caption{{\it Simulation of cells  which move according 
to $(\ref{velin1})$ when  the relaxation time $\tau_v(S)
\equiv \tau_v (S_t)$ is chosen so that $\tau_v$ is small 
(0.5 min) when $S_t$ 
is positive and large (5 min) when $S_t$
is not positive. The positions of
\/ $5000$ cells at different times are plotted.}}
\label{mod1fig}
\end{figure}

\smallskip
\noindent
{\bf (2)} In this case the relaxation time is specified as a function
of the time derivative of $S$ at the position of the cell, \ie, $\tau_v(S)
\equiv \tau_v (S_t)$. $\tau_v$  is  chosen so that
cells turn rapidly when the temporal derivative is positive and slowly
when it is negative. In our numerical example, we simply put $\tau_v=0.5$ 
min for $S_t > 0$, and $\tau_v = 10$ min for $S_t \le 0$. The results are 
shown in Figure \ref{mod1fig}; here one sees that the cells aggregate at the 
source of the waves. 

These cases show that reorientation that is adaptive with respect to
the temporal gradient of the signal suffices to produce aggregation,
as was found earlier in formal cell-based rules
\cite{Dallon:1997:DCM} and used previously in macroscopic descriptions
based on the classical chemotaxis equation \cite{Sperb:1979:MMD,Hofer:1997:SIS}.

Next we address the derivation of a macroscopic description from the
transport equation, using the direct effect of the signal on the
turning given by (\ref{velin1}). We denote by $p(\boldx,\boldv,t)$ the
density of individuals which are at point $\boldx \in \er^2$ and have 
velocity $\boldv \in V \subset \er^2$ at time $t$.  Here, $V$ is a bounded, 
symmetric set
which is determined by the external signal and by system
(\ref{velin1}). We also assume that there is a signal-independent
component to the turning for which the kernel $T$ is given by
$T(\boldv,\boldv') = (2 \pi v_0)^{-1} \delta(|\boldv -\boldv'| -v_0)$, 
where $v_0>0$ represents the magnitude of the random component of
$\boldv$. The cells add a small random component to their
velocity at a rate $\lambda$. Now $p(\boldx,\boldv,t)$
satisfies the transport  equation
\begin{align}
\label{startingequation}
\dfrac{\partial p}{\partial t} +   \nabla_\boldx \cdot \boldv p
+  \nabla_\boldv\cdot \left[\left(\dfrac{\boldg(\nabla S) 
- \boldv}{\tau_v(S)}\right)p 
\right]
= & \\ 
-\lambda p(\boldx,\boldv,t) + \frac{\lambda}{2 \pi v_0}& 
\int_V \delta(|\boldv -\boldv'| -v_0)
p(\boldx,\boldv'\!,t) \dboldv'. \nonumber
\end{align}
We  define the macroscopic density $n$ and
macroscopic flux $\boldj$  via 
\begin{equation}
n = \int_V p(\boldx,\boldv,t) \dboldv,
\qquad \qquad
\boldj = \int_V \boldv p(\boldx,\boldv,t) \dboldv,
\label{formulanj}
\end{equation}
and by integrating (\ref{startingequation}) over $\boldv,$ and multiplying
(\ref{startingequation}) by $\boldv$ and integrating over $\boldv,$ we obtain
the following evolution equations for $n$ and $\boldj$:
\begin{equation}
\dfrac{\partial n}{\partial t} +   \nabla_\boldx \cdot \boldj
= 
0
\label{equationforn}
\end{equation}
\begin{equation}
\dfrac{\partial \boldj}{\partial t} +   
\nabla_\boldx \cdot  \hat{\boldj} - \dfrac{1}{\tau_v(S)}\, n  \, \boldg(\nabla S)
= 
-  \dfrac{\boldj}{\tau_v(S)}.
\label{equationforj}
\end{equation}
The convective flux $\hat{j}_{ik} = \int_V v_i v_k p(\boldx,\boldv,t)
\dboldv$ that appears in (\ref{equationforj}) introduces a
higher-order moment, and in earlier work on bacterial chemotaxis we
could justify the closure hypothesis $j_{ik} = s^2 n \delta_{ik}/2$,
where $s$ is the speed of a bacterium (\cf
\cite{Erban:2005:STS}). Since the speed is not constant during
``runs'' in the amoeboid case, we must use a different approach
here. By constructing the evolution equation for $\hat{j}_{ik}$ and
assuming that it relaxes rapidly in time, \ie, neglecting its time
derivatives, and neglecting the third order velocity moments,
we find that the $2 \times 2$ tensor $\hat{j}$ has
components
\begin{equation}
\hat{j}_{ik}(\boldx,t) = 
\dfrac{\tau_v(S)\lambda v_0^2}{4} n \delta_{ik}
+
\dfrac{1}{2}\left(g_i j_k + g_k j_i \right),
\qquad \mbox{for} \; i, k = 1,2.
\label{jik}
\end{equation}
This leads to two possible closures, (i) by keeping only the
zero-order moment (the term involving n), or (ii) by keeping both the
zero-order and first-order contributions. We use the first of these
here and find that 
 the system
(\ref{equationforn}) -- (\ref{equationforj}) becomes 
\begin{equation}
\dfrac{\partial n}{\partial t} +   \nabla_\boldx \cdot \boldj
= 
0
\label{equationfornc}
\end{equation}
\begin{equation}
\dfrac{\partial \boldj}{\partial t} +
\nabla\left(\dfrac{\tau_v(S)\lambda v_0^2}{4} n(\boldx,t)
\right)  
- \dfrac{1}{\tau_v(S)} \, n  \, \boldg(\nabla  S)
= 
- \dfrac{\boldj}{\tau_v(S)}.
\label{equationforjc}
\end{equation}

\medskip\noindent
In this form we identify the chemotactic velocity and the chemotactic
sensitivity as 
$$
\boldu_c = \dfrac{1}{\tau_v(S)}  \boldg(\nabla S)
\qquad  \chi =  \dfrac{1}{\tau_v(S)}\dfrac{ \boldg(\nabla S)}{\nabla S } .
$$
where the latter only makes sense if we assume that $\boldg(\nabla S)
= g(\nabla S) \nabla S$.  If in addition $g$ saturates for large arguments, the 
velocity saturates and the sensitivity goes to zero in the presence of large
gradients, as one should expect. 

One sees that at this level of closure,
the relaxation rate of the flux on the right hand side of
(\ref{equationforjc}) is signal dependent, but if we were to suppose
that $\tau_v(S) \equiv \tau_0$ is independent of $S$ then the system
(\ref{equationfornc}) -- (\ref{equationforjc}) can be written as the
second order equation
\begin{equation}
\dfrac{\partial^2 n}{\partial t^2} +\dfrac{1}{\tau_0}\dfrac{\partial
n}{\partial t} 
=
\dfrac{\tau_0 \lambda v_0^2}{4} \triangle n- 
\nabla_\boldx \cdot \left( 
\dfrac{1}{\tau_0} \, n  \, \boldg(\nabla S)
\right),
\label{chemequation}
\end{equation}
which is the hyperbolic form of the classic chemotaxis equation.
However, if $\tau (S)$ is signal dependent the system
(\ref{equationfornc})--(\ref{equationforjc}) does not reduce to
(\ref{chemequation}), and this suggests that one cannot expect to
obtain the classical form of the chemotaxis equation when internal
states are taken into account explicitly. On the other hand, as we saw in the
simulations above, one cannot avoid the ``back-of-the wave paradox''
without a signal-dependent $\tau_v$
\cite{Dallon:1997:DCM,Dolak:2005:KMC}.
Let us note that we can treat similarly a modification of (\ref{velin1})
where we allow the force to
depend directly on the time derivative of the signal. This can be done
by replacing  $\boldg(\nabla S)$  with  $\boldg(\nabla S, S_t)$.

\begin{figure}
\MpicturesAB{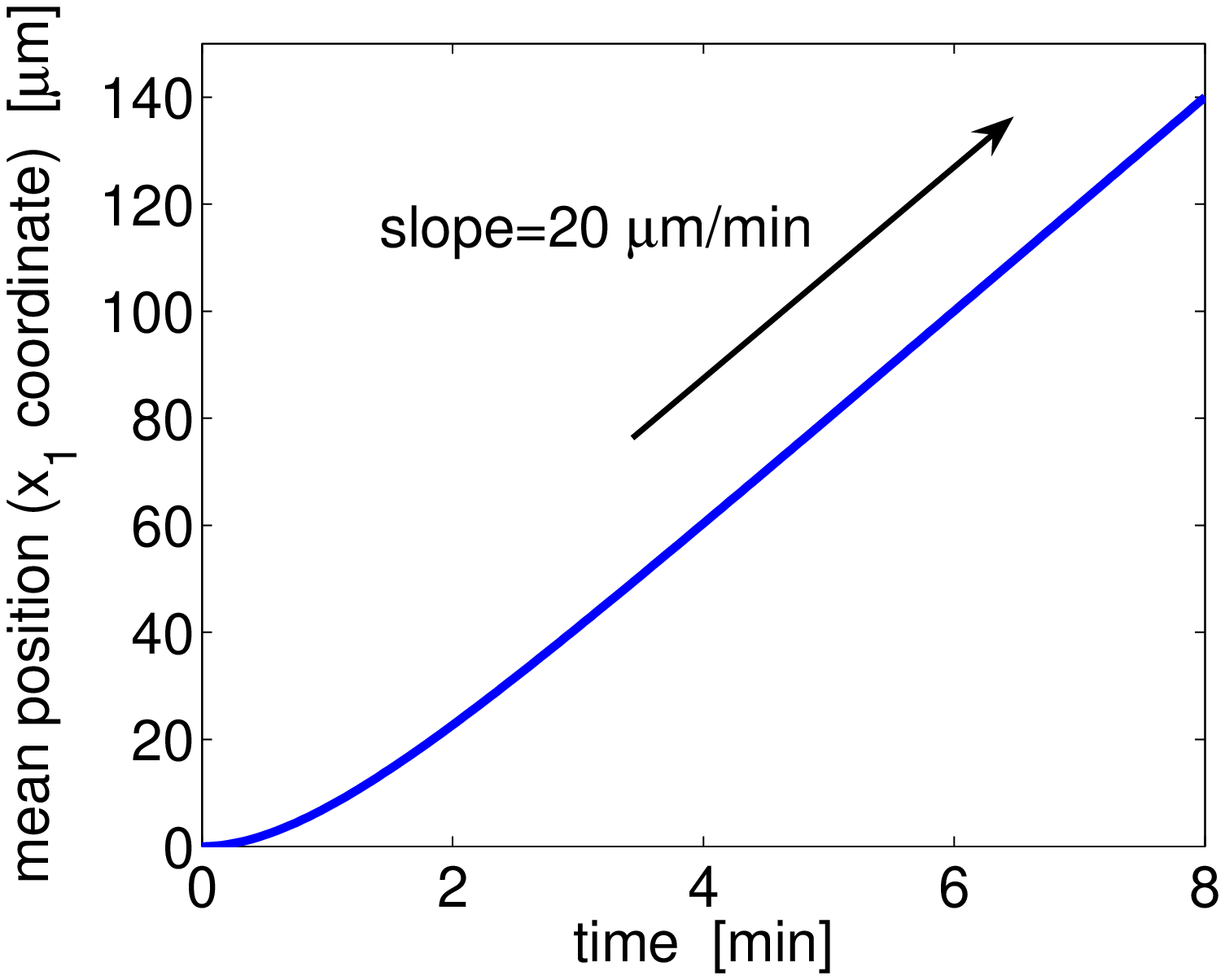}{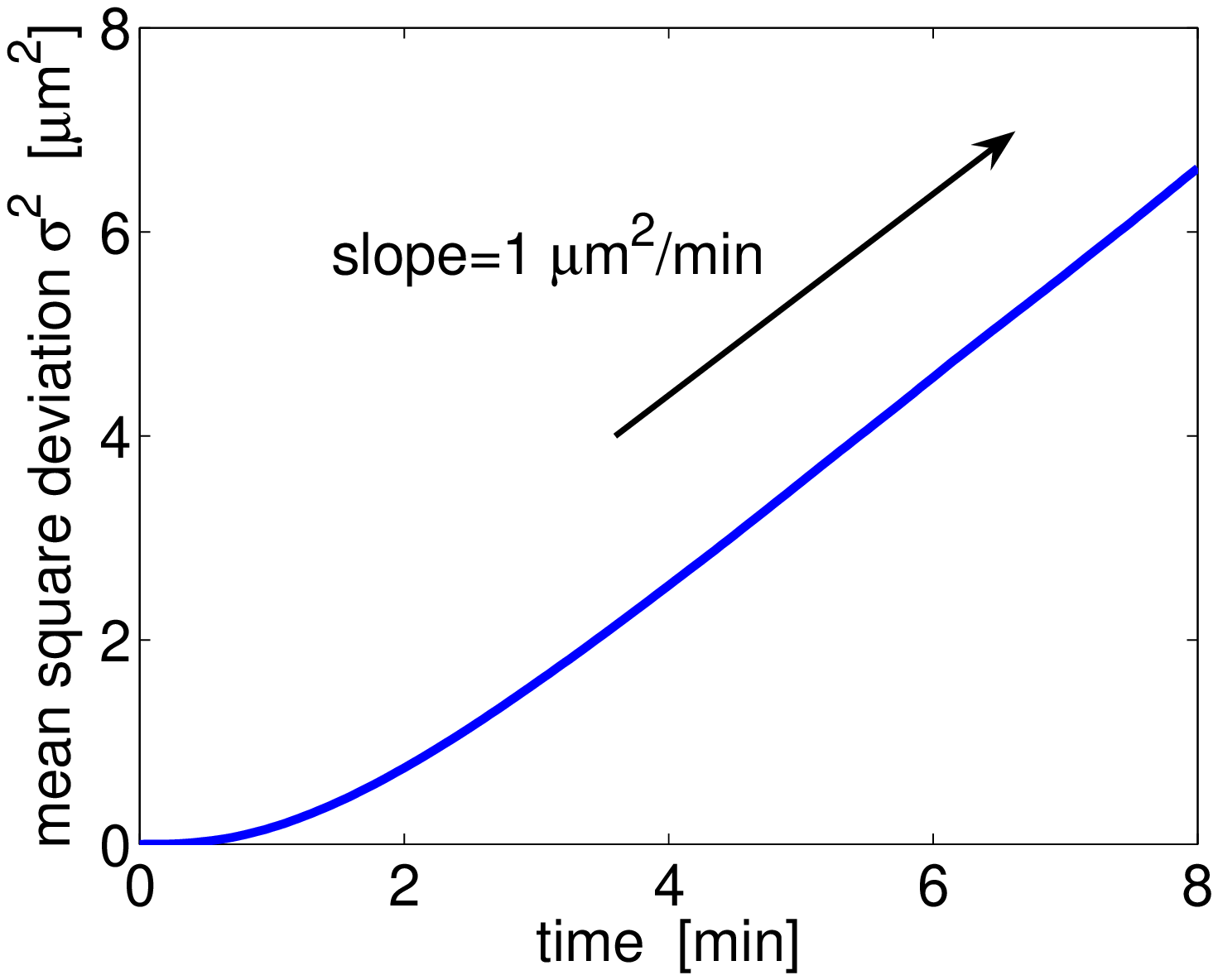}{1.5in}
\caption{{\it Time evolution of statistics $(\ref{datameanmsd})$
obtained from the cell-based simulation.}
(a) {\it First component of \/ $\boldX(t)$ as a function
of time.} (b) {\it Mean square displacement \/ $\sigma^{2}(t)$
as a function of time.}}
\label{figtestmodA} 
\end{figure}

To illustrate the validity of the macroscopic equation (\ref{chemequation}),
let us consider the cell-based numerical simulations of $N_0$ individuals
whose velocity is governed by  (\ref{velin1}). We denote the positions
of individuals as $\boldx_i(t)$, $i=1,\dots,N_0$. Then the quantities
of interest are the mean position of individuals and the mean square deviation,
and for the discrete-cell analysis these are defined as
\begin{equation}
\boldX(t) 
=
\frac{1}{N_0} \sum_{i=1}^{N_0} \boldx_i(t)
\qquad
\mbox{and}
\qquad
\sigma^{2}(t)
=
\frac{1}{N_0}
\sum_{i=1}^{N_0} 
\parallel \!\!\boldx_i(t) - {\boldX}(t)\!\!\parallel^{2}.
\label{datameanmsd}
\end{equation}
By multiplying  (\ref{chemequation})  by $\boldx$ and integrating over
$\boldx$, then computing the variance, much as in Section 1.1, we find
that for long times the macroscopic description predicts that 
\begin{equation}
\boldX(t)
\approx
\boldg(\nabla S) \, t
\qquad
\mbox{and}
\qquad
\sigma^{2}(t)
\approx
\tau_0^2 \lambda v_0^2 \,\! t.
\label{estimatedatameanmsd}
\end{equation}
To compare the theoretically-derived results
(\ref{estimatedatameanmsd}) with the cell-based computations, we
choose $\tau_0 = 1$ min, $\lambda= 1 \; \mbox{min}^{-1}$, $v_0=1
\; \mu\mbox{m}/\mbox{min}$, $N_0=10^4$ cells, $\boldg=\omega {\mathbf I
\mathbf d}$, where $\omega = 20 \; \mu\mbox{m}^2/$min, and $\nabla S$ is
 given by (\ref{testgradient}) with $\norm{\!\!\nabla S \!\!}=1
\; \mu\mbox{m}^{-1}$. We place all cells at $[0,0]$ and set their
velocities to 0 initially, compute their subsequent motion, and plot
the first component of $\boldX(t)$ and $\sigma^{2}(t)$ (as given by
(\ref{datameanmsd})) in Figure
\ref{figtestmodA}.
We see that after an initial transient period both quantities grow
linearly with time, and the slopes are asymptotically equal to the
slopes predicted theoretically using (\ref{estimatedatameanmsd}).

\subsection{The infinite-dimensional model and its finite-dimensional
reduction
}

\label{seccartooneuk}

As discussed  earlier,  analysis of {\it E.coli} chemotaxis shows that the
microscopic behavior can be translated into the macroscopic
parameters, and it is desirable to do the same for amoeboid
chemotaxis. However, as noted earlier the internal state may now live
in a Banach space, and a reduction to finite dimensions is
necessary. We start with the description of excitation-adaptation
dynamics on the cellular membrane to model directional sensing and
reduce the resulting system.  For simplicity we suppose that a cell is
a disk of radius $d.$ The state of a cell will be described by the
position $\boldx$ and velocity $\boldv$ of its centroid, and several internal
variables on the membrane. The membrane of the cell can be described
as the set 
\begin{equation}
M =  \big\{d \, [\cos(\theta),\sin(\theta)] \,\, | \,\, \theta \in [0,2 \pi) 
\big\}.
\label{membrane}
\end{equation}
The local state at each point of the membrane will be specified by
the (infinite-dimensional) internal state variable 
$~~\boldy(\theta, t) \equiv [y_1(\theta,t),y_2(\theta,t)]^T$, $\theta
\in [0,2 \pi)$, whose  evolution is
governed by the ``excitation-adaptation" cartoon model
(\ref{rom81}). In the formalism of equations (\ref{intradyn1}) --
(\ref{intradyn2}) this means that the internal state $\boldy(t) :
[0,2\pi) \to \er^2$ can be viewed as an element of the Banach space of
$2\pi$-periodic vector functions $\mathbb Y$ and evolves according to
\begin{equation}
\dfrac{\partial \boldy}{\partial t} (\theta,t)
= \tS(\theta,t) {\boldtau} - \cT \boldy(\theta,t)
\label{y1eqtheta}
\end{equation}
for $\theta \in [0,2 \pi)$ and $ t > 0$.  Here 
$$
\tS(\theta,t) = S \big( \boldx + d~\et(\theta),t), \quad  \quad
\et(\theta) = [\cos \theta, \sin \theta]^T,\quad  \quad
{\boldtau} = \left[ \tau_e^{-1},\tau_a^{-1} \right]^T,
$$
and
$$
\cT \equiv \left[\begin{array}{cc}
\tau_e^{-1} & \tau_e^{-1} \\
0 & \tau_a^{-1} \end{array}\right].
$$
The $\boldy$-variables  correspond 
to those in (\ref{intradyn1}), and we project these to finite
dimensions by considering the first Fourier mode of $y_1$ and the
first two Fourier modes of $y_2$. Thus we define 
 the average internal variables as
\begin{eqnarray}
\label{zdef}
\boldz(t) 
=  \left( z_1(t),z_2(t)\right)^T &=& 
\dfrac{1}{2 \pi} \int_0^{2 \pi} \boldy(\theta,t) d \theta, \\
\boldq(t) 
=  \left( q_1(t),q_2(t)	\right)^T &=& 
\dfrac{1}{d \pi} \int_0^{2 \pi} \et(\theta)  y_2(\theta,t) d \theta.
\label{q1q2def}
\end{eqnarray}
To derive equations for the reduced finite-dimensional
set of internal variables $\boldz(t)$ and $\boldq(t)$,
we use the approximation 
\begin{equation}
\tS(\theta,t) \sim S(\boldx,t) + 
d~ \et(\theta) \cdot \nabla S,
\label{sigapprox}
\end{equation}
and consequently, we can write 
\begin{equation}
\dfrac{\partial y_2}{\partial t} (\theta,t)
= \dfrac{S(\boldx,t) + d~ \et(\theta) \cdot \nabla S - y_2 (\theta,t)}{\tau_a}.
\label{y2eqthetab}
\end{equation}
for $ \theta \in [0,2 \pi)$ , $ t > 0$. 
Multiplying (\ref{y2eqthetab}) by 1, $\cos(\theta)$ 
or $\sin(\theta)$ and integrating the resulting equations
with respect to $\theta,$ we obtain
\begin{eqnarray}
\label{z2eq}
\dfrac{\dz_2}{\dt} &=& \dfrac{S(\boldx,t) - z_2}{\tau_a}\\
\dfrac{\dboldq}{\dt}
&=& \dfrac{\nabla S(\boldx,t) - \boldq}{\tau_a},
\label{bqeq}
\end{eqnarray}
Thus $z_2$ relaxes to the signal $S(\boldx,t)$ and $\boldq$ relaxes to
the directional information of $S$, both with the decay rate $\tau_a$.
To interpret $z_1$, we assume fast excitation (\ie, $\tau_e = 0).$
Then using the fact that $y_1 = \tS - y_2$ and integrating
(\ref{y2eqthetab}) with respect to $\theta$, one finds that
\begin{equation}
\dfrac{\dz_1}{\dt}
= \dfrac{\partial S}{\partial t}  (\boldx,t)
+ \boldv \cdot \nabla S(\boldx,t) - 
\dfrac{z_1}{\tau_a},
\label{z1eqt0}
\end{equation}
By integrating  this one  sees that $z_1$ tracks the Lagrangian derivative of
$S$ taken along the cell's trajectory, with a memory  determined by
$\tau_a$: the smaller $\tau_a$ the faster the cell forgets the history
of this derivative. Taken together, the four variables $ (z_1, z_2, q_1, q_2)$
contain information about the rate of change of the signal along the
trajectory ($z_1$), the local value of the signal ($z_2$), and the
gradient of the signal ($\boldq$). To this set we will add the
polarization axis $\boldu = (u_1,u_2)$ in the next section, and the
result will be the smallest set of variables that is able to capture
the phenomena described earlier.  Consequently, the simplest
hypothesis is that cellular motility depends only on these six
variables, \ie, the $\boldz$ used in (\ref{zequation}) has six
components.

\subsection{The  motility model }

\label{sectionincoeukar}

 The next step is to build this model for the internal dynamics into a
 description of cellular movement in order to reproduce some of the
 experimental behaviors observed for eukaryotic chemotaxis. As we saw
 in Section \ref{secdirection}, Dd or leukocytes
 respond to the waves of chemoattractant by moving toward the source
 of waves, and the five different stages of the wave with different
 behavioral responses of the cell are schematically shown in Figure
 \ref{figwave}. These and various other cell types also polarize after
 sufficient exposure to a directional signal
 \cite{Iijima:2002:TSR}. In order to build directional sensing,
 polarization and response to waves into the model, we distinguish
 three distinct states of cells: (1) polarized cells which are motile
 (MPC), (2) polarized cells which are resting (\ie, non-motile,
 denoted RPC), and (3) non-polarized cells which are resting (RUC).
 The signal transduction machinery for all types is described by the
 membrane-based model (\ref{membrane}) -- (\ref{y1eqtheta}); 
 the difference between the
 types is in their motility behavior.

We describe a motile polarized cell (an MPC) by its position $\boldx$,
its velocity $\boldv$ and its internal state $\boldy$, as
before. However, instead of defining the force directly in terms of
the signal, as was done in (\ref{velin1}), we assume that the force is
proportional to the projected internal variable $\boldq$
(defined by (\ref{q1q2def})), which tracks
the gradient of the signal (\cf (\ref{bqeq})). Thus we write the
equations of motion for a cell as
\begin{equation}
\dfrac{\dboldx}{\dt}  = \boldv, \qquad \dfrac{\dboldv}{\dt}  =
\dfrac{\gamma \boldq - \boldv}{\tau_d}.
\label{movvelocity}
\end{equation}
 In a steady gradient of the signal $\boldq$ relaxes to
$\nabla S$ on the time-scale $\tau_a$, and $\boldv$ relaxes to $\gamma
\boldq$ on the time-scale $\tau_d$; thus the models predicts steady
motion in a constant gradient. One expects that in general $
\tau_a < \tau_d$. 
However, as we saw earlier, to explain the back-of-the-wave behavior
\cite{Geiger:2003:HPL} the response to the wave must be biased toward
moving when the signal is increasing in time, as in the front of the
wave.  It is known that Dd cells and leukocytes stop translocating and lose
their polarity in Phase (C) of a wave (\cf Figure \ref{figwave}), which
introduces an asymmetry into the response, and to capture this we
introduce a resting state.  A resting cell (either an RPC or an RUC)
is described by its position $\boldx$ and its internal state $\boldy
\in \mathbb Y$, and these cells may  also have a polarization
axis $\boldu = (u_1,u_2)$.  We assume that the position of a resting
cell is fixed and
that the internal state evolves according to (\ref{y1eqtheta}).

\begin{figure}
\centerline{\psfig{file=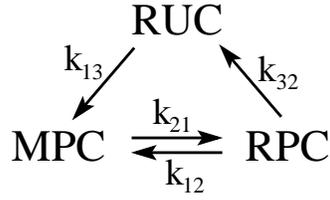,height=1.in}}
\caption{{\it The allowed transitions between cell states. The transition
rates depend on the internal state $z_1$ as follows: 
$k_{21} = \lambda_1 - b_1z_1$, 
$k_{12} = \lambda_2 + b_2z_1$, 
$k_{13} = \lambda_3 + b_3z_1$, 
$k_{32} = \lambda_0$.}}
\label{figrates} 
\end{figure}

 Finally, we must postulate how transitions between the three states
depend on the signal (\cf Figure \ref{figrates}).  We assume that the
motile cells retain their polarity upon stopping, and that the
transition rate from the motile to the resting polarized state depends
on $z_1$ as shown in Figure \ref{figrates}. A moving cell ``computes''
the directional vector $\boldq$ and the average around the perimeter
of the internal variable $y_1$, which is $z_1$, according to
(\ref{bqeq}) and (\ref{z1eqt0}).
The interpretation of Figure \ref{figrates} and the justification for
the postulated dependence of the transition rates between states on
$z_1$ are as follows.

\medskip
\noindent
{\bf (i)} If the Lagrangian derivative of the signal along a cell's
trajectory is negative, then $z_1$ decreases and $k_{21}$, the
transition rate from the motile state to the resting polarized state,
increases.  The resting polarized cell adopts a polarization equal to
the velocity vector before it stops, \ie, $\boldu= \boldv$ after the
transition.

\medskip
\noindent
{\bf (ii)} If a cell is resting and there is an increase in the
signal, then $z_1$ increases and it is more 
likely to move. It the cell is unpolarized, than its initial 
velocity (polarization) is zero and the time $\tau_d$ reflects the time
delay needed for polarization of the cell.
If the cell was already polarized than its initial
velocity is equal to the polarization vector, \ie,. 
we set $\boldv=\boldu$,  and the time delay $\tau_d$ reflects
 the relaxation time for turning, if it is necessary.

\medskip
\noindent
{\bf (iii)} A resting polarized cell looses its polarity at a rate
$\lambda_0,$ and thus polarized cells which do not receive a stimulus
for a long time lose their polarity.

\medskip
\noindent
Next we demonstrate that 
the model can successfully solve the back-of-the-wave
problem \cite{Dolak:2005:KMC,Geiger:2003:HPL}, \ie,
cells will aggregate at the source of the attractant waves.

\subsection{Aggregation when resting states are incorporated}

\label{secnumsim}

The internal dynamics model $\boldy$ written in terms of (\ref{intradyn1})
is given by (\ref{membrane}) -- (\ref{y1eqtheta}), and every cell is
described by its position $\boldx \in \er^2$, its velocity
$\boldv \in \er^2$, its polarization axis $\boldu \in \er^2$ and its
internal state function $\boldy \in \mathbb Y$.  To compute $z_1$ and
$\boldq$, the radius of a cell is set to $d=7.5\;\mu$m, and we
discretize the cell boundary (\ref{membrane}) using $m$ meshpoints,
$$
\theta_j = \frac{2 \pi j}{m},
\qquad
\mbox{for} \quad j = 1, 2, \dots, m-1,m.  
$$
Then the state of each cell is described by an $(m+4)$-dimensional
 vector
\begin{equation}
(\boldx, \boldv,y(\theta_1),y(\theta_2), \dots, y(\theta_m)).
\label{discrstate} 
\end{equation}
Here, $\boldv$ denotes the velocity for an MPC and  the polarization axis
for an RPC.  We can simply set this  equal to 0 for RUCs, which are in
fact described by $m+2$ variables.  The internal state variables
$y(\theta_j)$ evolve according to $m$ equations of
the form (\ref{y1eqtheta}), which are uncoupled because there is no
transport along the membrane. The evolution of $\boldx$ and $\boldv$
is described in Section \ref{sectionincoeukar}. At each time step we
use the $y(\theta_j)$ to numerically approximate
integrals (\ref{zdef}) -- (\ref{q1q2def}) and thereby  compute $z_1$,
which is needed for the  computation of the 
transition rates between different states as shown in Figure
\ref{figrates},  and $\boldq$, which is necessary for the integration of
(\ref{movvelocity}).  Throughout we use $ m = 50$, and therefore every
cell is described by a $54-$dimensional state vector
(\ref{discrstate}).

\begin{figure}[hbt!]
\centerline{
\psfig{file=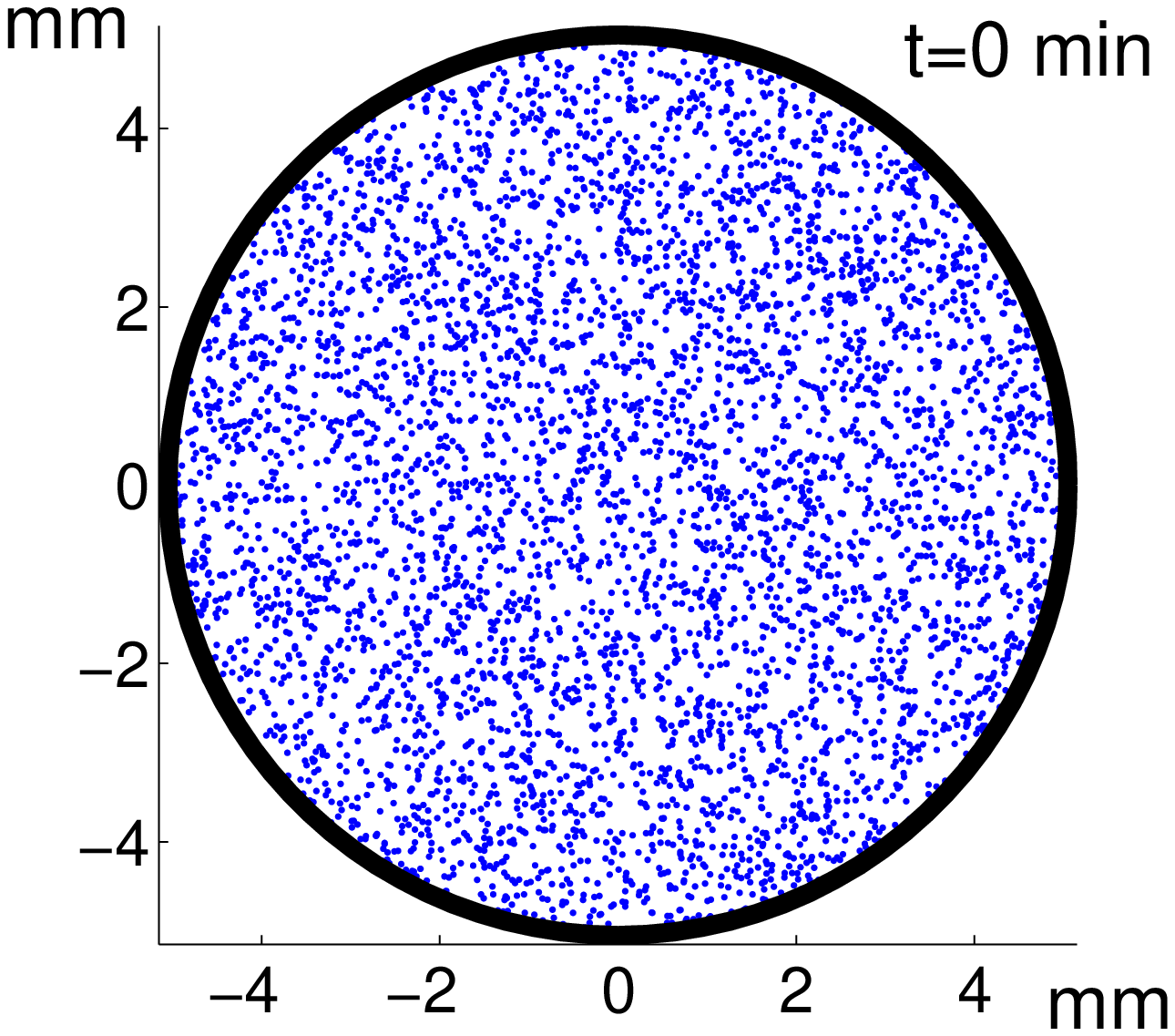,width=2.25in}
\psfig{file=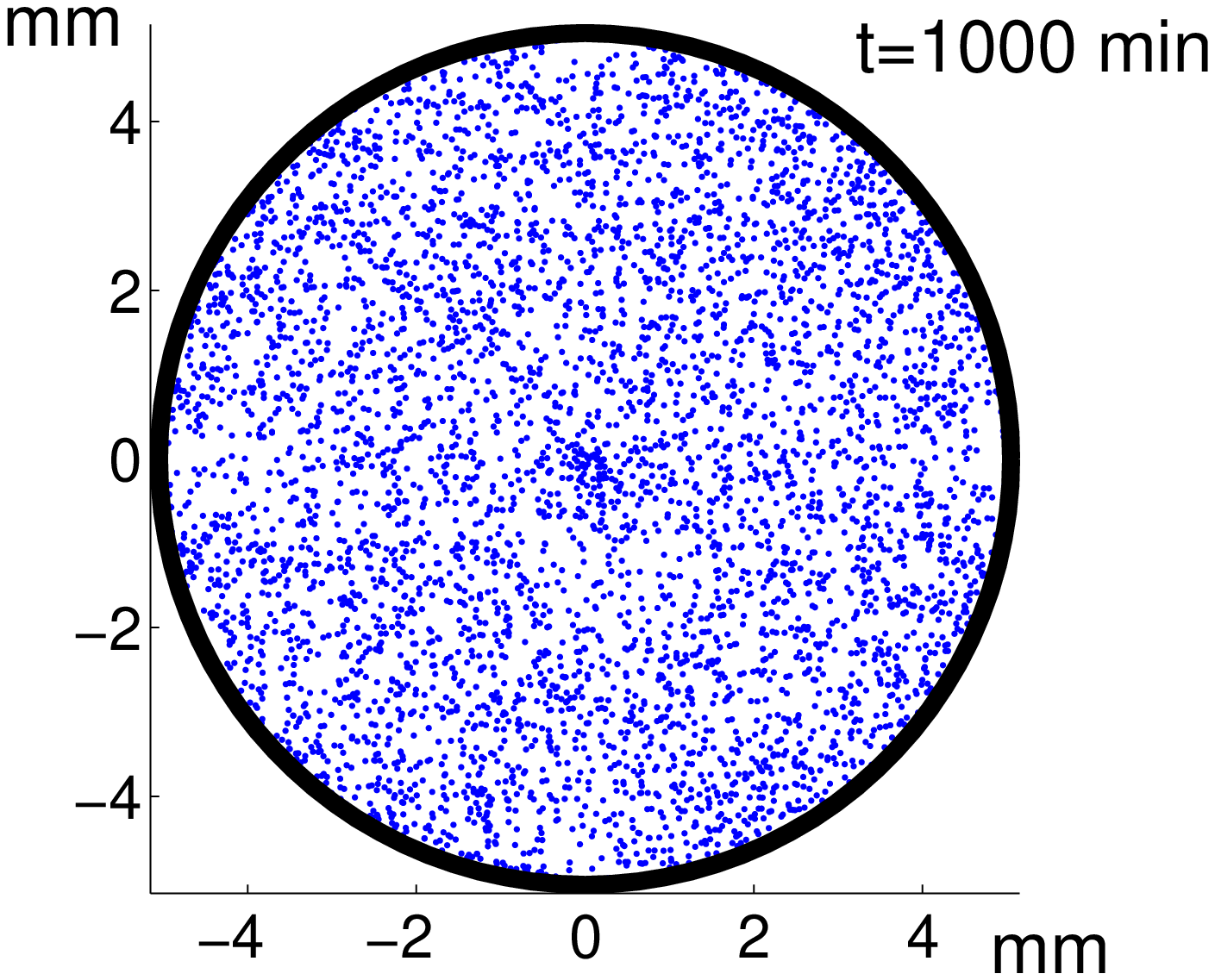,width=2.25in}
}
\caption{{\it The cell distribution as a function of time for $b=0$. 
Positions of $5000$ cells at times $0$ min  and $1000$ min
for periodic waves of chemoattractant.}}
\label{figintb0} 
\end{figure}

\begin{figure}[hbt!]
\centerline{
\psfig{file=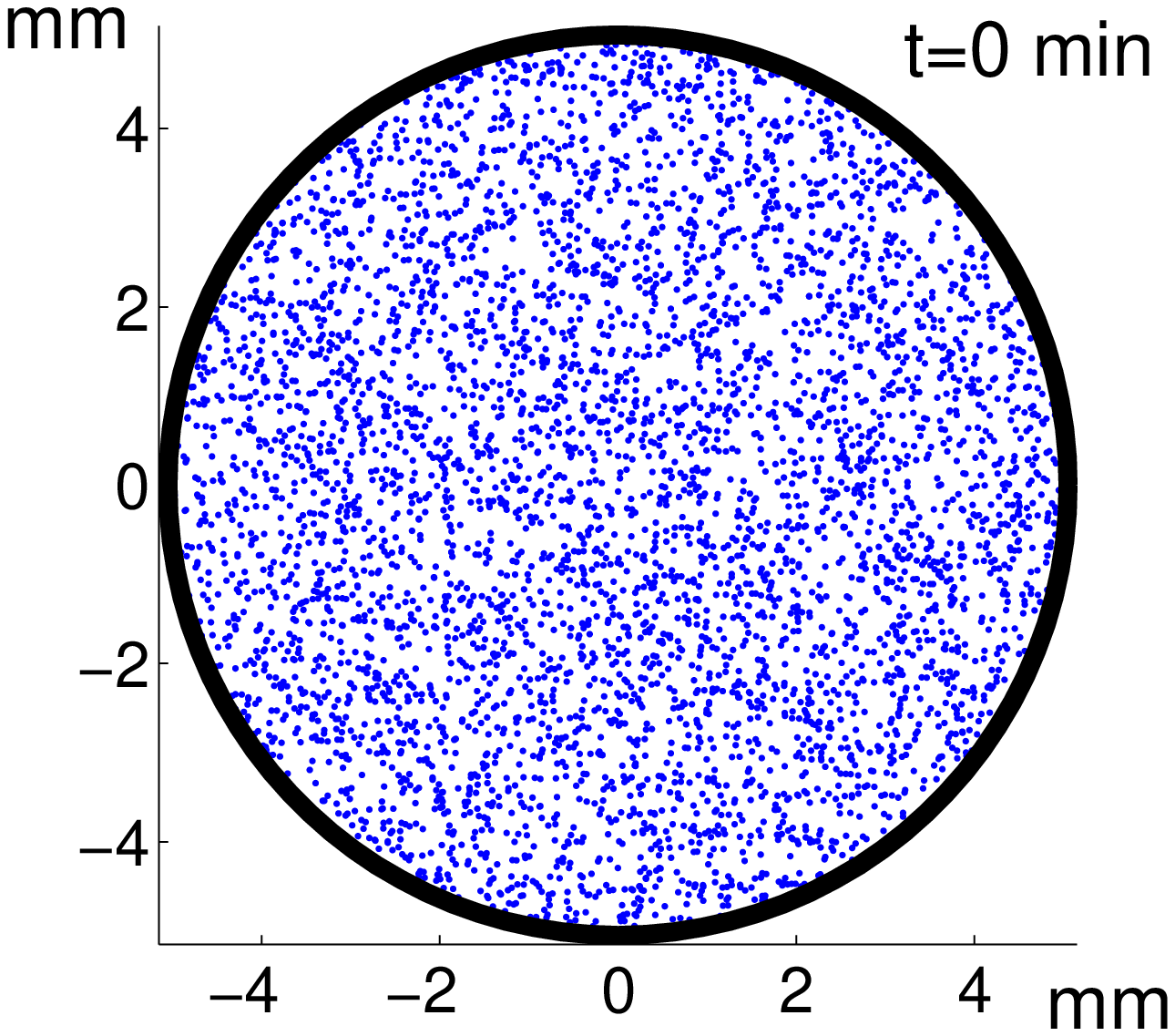,width=2.25in}   
\psfig{file=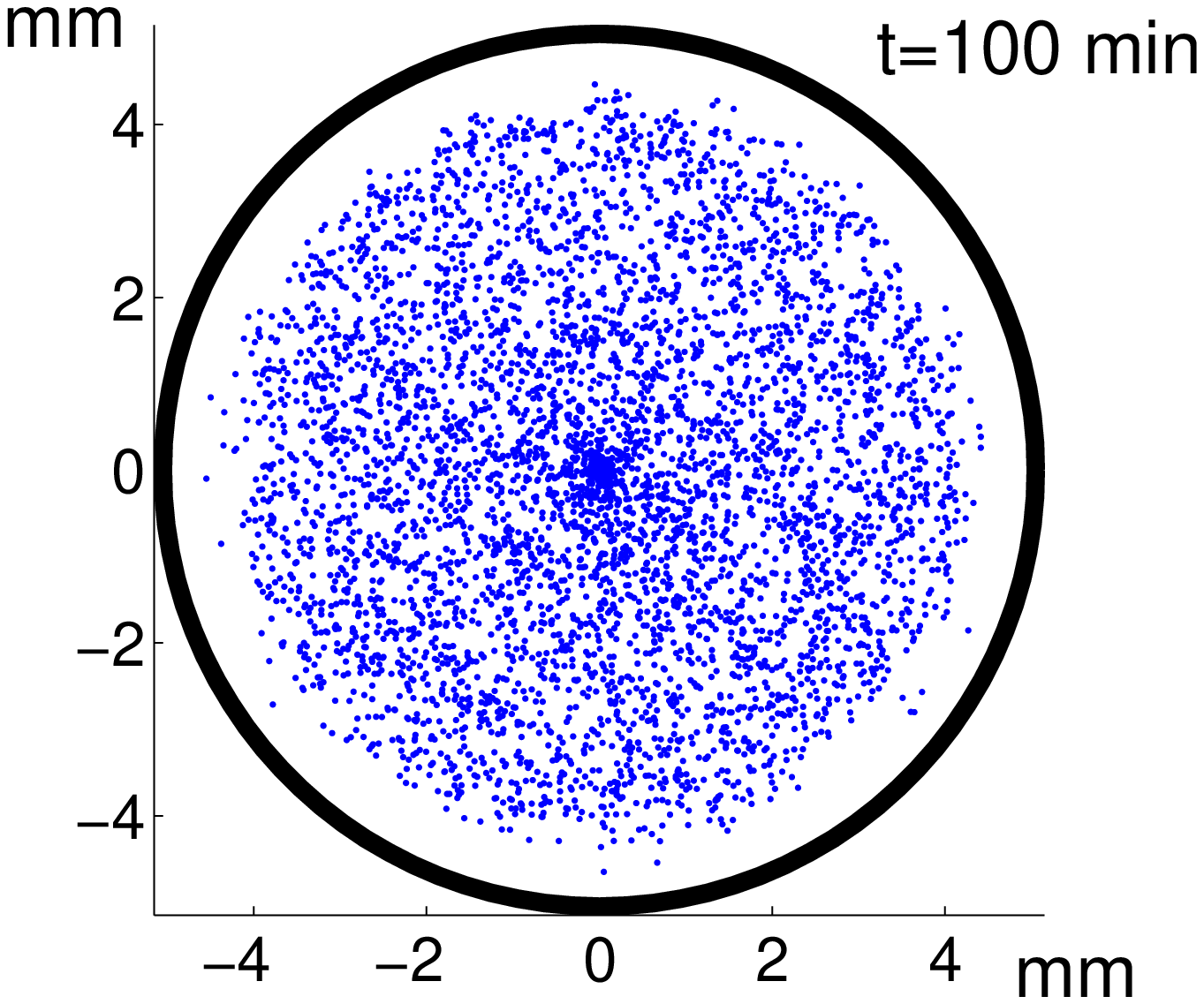,width=2.25in}
}
\centerline{
\psfig{file=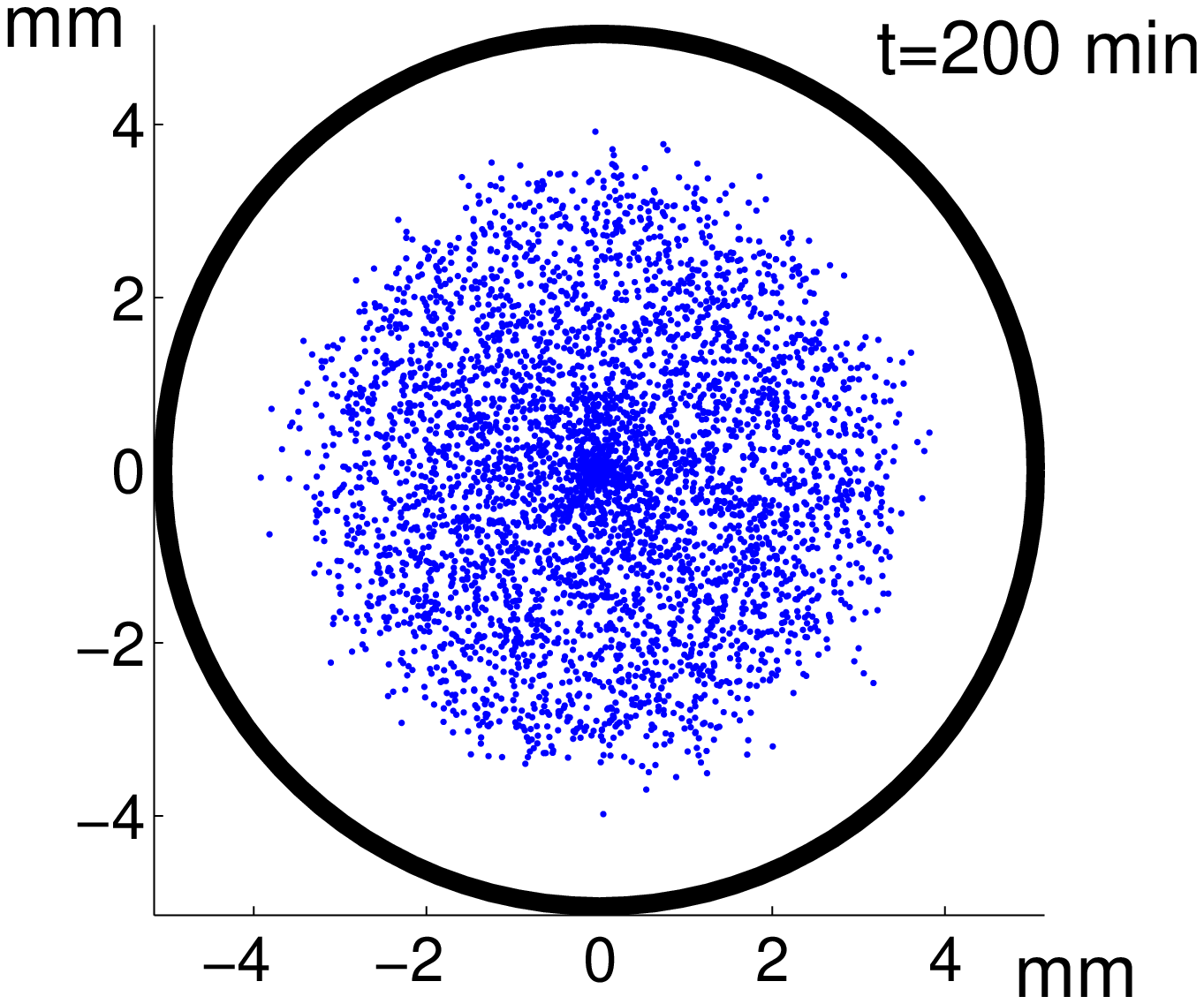,width=2.25in}
\psfig{file=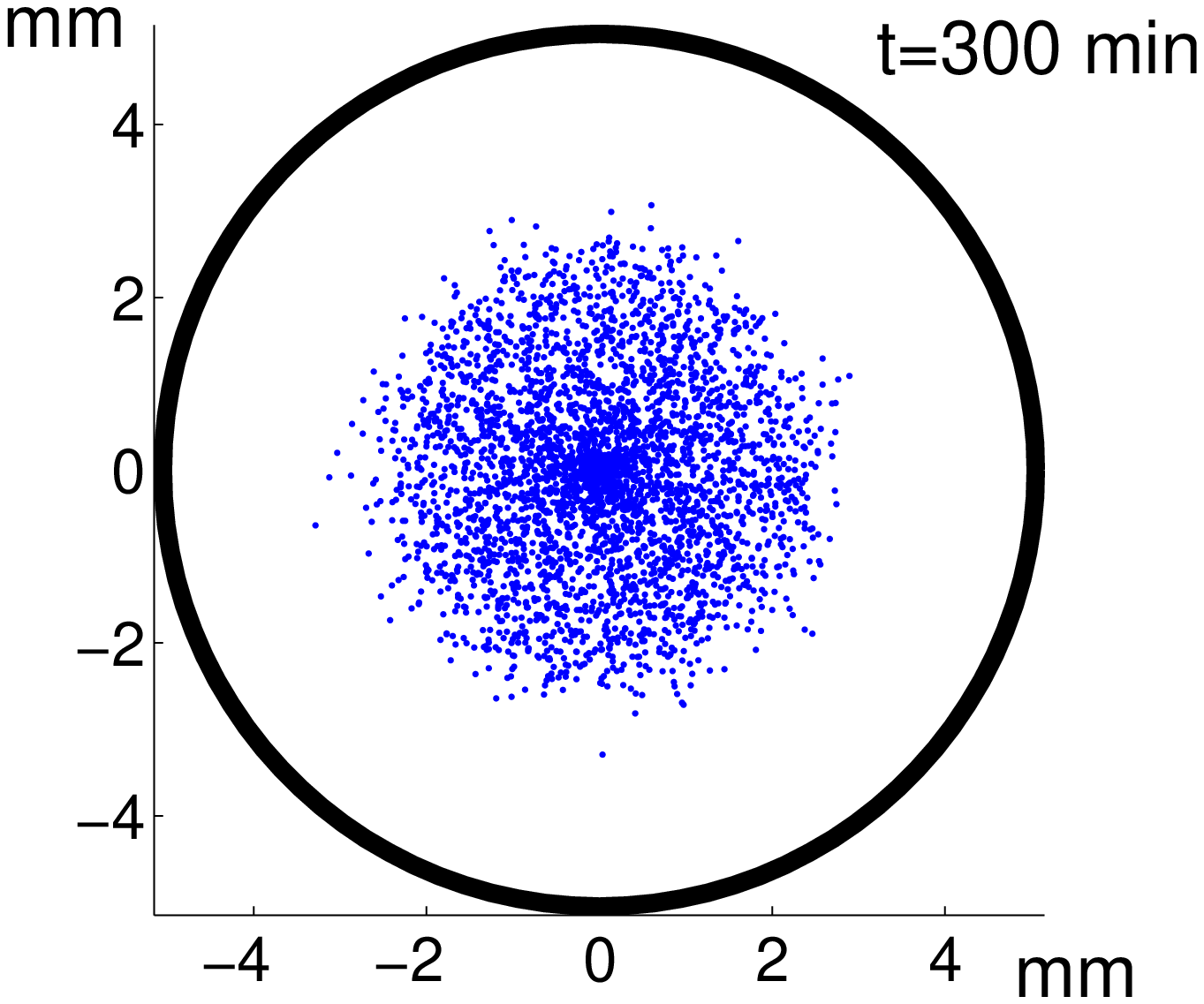,width=2.25in} 
}
\centerline{
\psfig{file=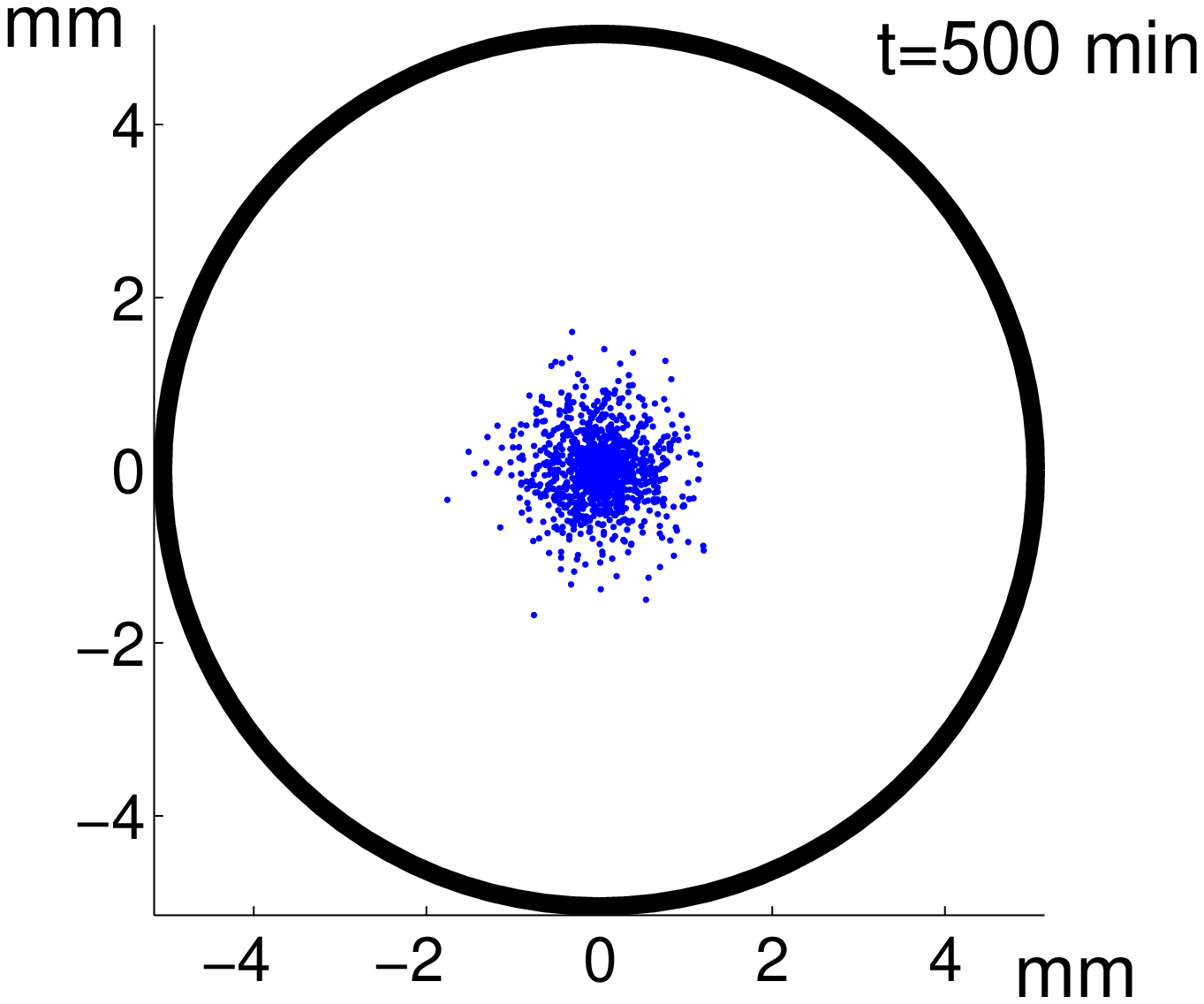,width=2.25in}
\psfig{file=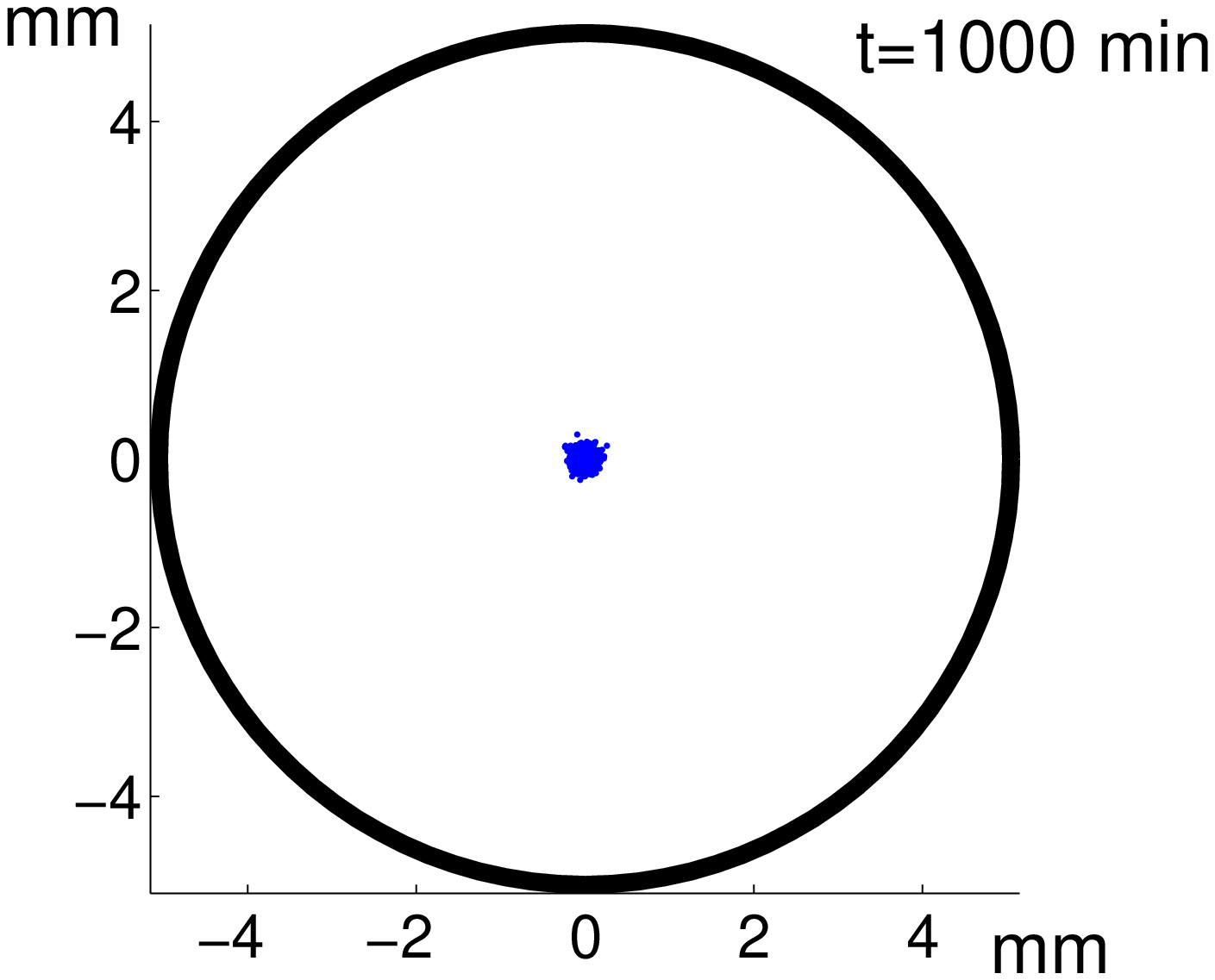,width=2.25in}
}
\caption{{\it The cell distribution as a function of time for  
$b=2\; \mbox{min}^{-1}$.  
Positions of $5000$ cells at times $0$ min, $100$ min, $200$ min, $300$ min, 
$500$ min and $1000$ min for periodic waves of chemoattractant.}}
\label{figintb2} 
\end{figure}

As was done for Figures \ref{mod1figB} and
\ref{mod1fig}, we consider a two-dimensional disk  of
radius $5$ mm, and we specify a periodic source of cAMP waves at the
center of the domain.  The period of the waves is seven minutes, their
speed is $400\;\mu{}\mbox{m/min}$, and the waves are scaled so that the
maximum $\norm{\!\!\nabla S\!\!}$ is 1 mm$^{-1}$.  Initially the cells are
distributed uniformly.  We use the following base transition rates and
sensitivities in the transition rates $k_{ij}$ given in Figure \ref{figrates}:
\begin{equation}
\lambda_1 = \lambda_2 = \lambda_3 = 1 \; \mbox{min}^{-1},
\quad
\lambda_0 = 0.2 \; \mbox{min}^{-1}
\quad
\mbox{and}
\quad
b_1 = b_2 = b_3 \equiv b
\label{tranparam} 
\end{equation}
Later the parameter $b$  will be varied (\cf Figures \ref{figintb0}
 and \ref{figintb2}).  The time constants are chosen as
\begin{equation}
\tau_e = 0 \quad \mbox{(fast excitation)},
\qquad
\tau_a = 0.5 \; \mbox{min},
\qquad
\tau_d = 2 \; \mbox{min}.
\label{timeparam} 
\end{equation}
The  two parameters which have yet to be specified are $\gamma$ in
equation (\ref{movvelocity}) and $b$. The parameter $\gamma$ simply
rescales the speed of cells. We know from experiments that the maximal
speed of a cell is about 20 $\mu$m per minute, which can be used to
fit the value of the parameter $\gamma$. We found that for
$\gamma=0.08$ mm$^2$/min, the average speed of cells on the steepest part of
the wave front is between 10 $\mu$m per minute and 20 $\mu$m per
minute in all simulations.  Hence, we used $\gamma=0.08$ mm$^2$/min to compute
the plots shown in Figures \ref{figintb0} and
\ref{figintb2}.

The parameter $b$ specifies how strongly the turning rates depend on
$z_1$ and we tested three possibilities $b=0$, $b=1\; \mbox{min}^{-1}$
and $b=2\; \mbox{min}^{-1}.$ If $b=0$ the transition rates $k_{ij}$
are independent of $z_1$, and the time evolution of the cell positions
is shown in Figure \ref{figintb0}. We see that in this case there is
no aggregation, which is similar to what was shown earlier in Figure
\ref{mod1figB} where we considered the model without the internal
dynamics and with a constant relaxation time.
The computational results for 
$b=2\; \mbox{min}^{-1}$ are shown in Figure \ref{figintb2}, where the 
aggregation time is comparable to
the eight hours observed experimentally.   The results for $b =
1\; \mbox{min}^{-1}$ are similar - the only difference is that the aggregation is slower
(results not shown).  
Using (\ref{tranparam}), we see that the transition
rates (which depend on $z_1$) can be expressed in units of min$^{-1}$ 
as 
$  
k_{12} = k_{13} = 1 + b z_1
$
and
$
k_{21} = 1 - b z_1.
$ 
Since $b z_1$ is approximately in range $[-0.35,0.35]$ min$^{-1}$ for 
$b = 1$ min$^{-1}$ and
in the interval $[-0.7,0.7]$ min$^{-1}$ for $b=2$ min$^{-1}$, it implies 
that the turning
rates are in the interval $[1-0.35,1+0.35]$ min$^{-1}$ for $b = 1$ min$^{-1}$
and in the interval $[1-0.7,1+0.7]$ min$^{-1}$ for $b=2$ min$^{-1}$.
As will be seen in Section \ref{transport}, the  moment approach used
there   is justified when   $b z_1$ is small,
\ie, for small bias of the turning rates. 
The error may increase significantly for large $b$ because the higher
order moments may not be negligible.

In these figures, as in the preceding ones related to aggregation,
there is no stream formation such as is observed during aggregation of
Dd. Here cells always move radially inward toward the source because
the waves are imposed and are axisymmetric. In the presence of signal
relay, as in Dd, it was shown earlier \cite{Lilly:2000:PFC} that
signal relay combined with a random initial distribution of cells
plays an essential part in stream formation.

\section{Transport equations }
\label{transport}

Next we  show that the microscopic model for signal detection,
transduction and movement can be embedded in a system of transport
equations and thence into a system of moment equations for macroscopic
quantities. To that end, note that every cell with the same
$\boldx,\boldv,z_1,\boldq$ and same polarization state will follow the
same rules for movement, so it is natural to introduce density
functions $p^1,p^2,p^3$ as follows:

\begin{itemize}

\item[$\bullet$] $p^1(\boldx,\boldv,z_1,\boldq)$ is the  density of moving
cells at position $\boldx$ with velocity $\boldv$
and internal moments $z_1$, $\boldq$;

\item[$\bullet$]  $p^2(\boldx,\boldu,z_1,\boldq)$ is the  density of resting
polarized cells at position $\boldx$ with polarization axis $\boldu$
and internal moments $z_1$, $\boldq$. To simplify
the form of resulting transport equations, we denote the polarization 
axis as $\boldu \equiv \boldv$ in what follows;

\item[$\bullet$]  $p^3(\boldx,z_1,\boldq)$ is the density of resting 
unpolarized cells at
position $\boldx$ and with internal moments $z_1$,  $\boldq$.
\end{itemize}
Hereafter we  assume  excitation is fast, \ie, $\tau_e = 0, $ and
we use the  approximation given at (\ref{sigapprox}) for the
signal. The evolution of internal variables $\boldq$, $z_1$
is therefore given by (\ref{bqeq}) -- (\ref{z1eqt0}).
In order to simplify the following equations for $p^1,$ $p^2$ and $p^3$,
we define an operator $\cL$ by
\begin{equation}
\cL r = \dfrac{\partial r}{\partial t} 
+ 
\nabla_\boldq \cdot 
\left[
\dfrac{1}{\tau_a}
\left(
\nabla S - \boldq \right)r
\right]
+
\dfrac{\partial}{\partial z_1} 
\left[
\left( \dfrac{\partial S}{\partial t}
- 
\dfrac{z_1}{\tau_a}
\right)
r 
\right],
\label{lrequation}
\end{equation}
then the transport equations for $p^1,$ $p^2$ and $p^3$ are
\begin{eqnarray}
\cL p^1 
&=& -\nabla_\boldx \cdot v p^1
-  
\nabla_\boldv \cdot \left[
\dfrac{1}{\tau_d}
\left(
\gamma \boldq - \boldv \right) p^1 
\right]
-
\dfrac{\partial}{\partial z_1} 
\left[
\left(
\boldv \cdot \nabla S
\right)
p^1 
\right]
\nonumber \\ [8pt]
\label{r1eq}
&& - (\lambda_1 - b_1 z_1) p^1 + 
(\lambda_2 + b_2 z_1) p^2 +  \delta_{\boldv}
(\lambda_3 + b_3 z_1)
p^3,  \\ [8pt]
\label{r2eq}
\cL p^2 
& = &
  (\lambda_1 - b_1 z_1) p^1 
- (\lambda_2 + b_2 z_1)p^2 
- \lambda_0 p^2, 
\label{r2equation} \\[8pt]
\label{r3eq}
\cL p^3
& = &
\lambda_0 \int_V p^2 \dboldv
-
(\lambda_3 + b_3 z_1)
p^3. 
\end{eqnarray}
where $\delta_{\boldv}$ is the Dirac function. 
Next we define the macroscopic densities of particles in different
states as follows: 
\begin{align}
\label{momn12}
n^1(\boldx,t) &  = 
\displaystyle \int_V \displaystyle \int_{\er^3} 
p^1(\boldx,\boldv,z_1,\boldq)
\, \dboldv \dz_1 \dboldq, \\[5pt]
n^2 (\boldx,t) &  =  
\displaystyle \int_V \displaystyle \int_{\er^3} 
p^2(\boldx,\boldv,z_1,\boldq)
\, \dboldv \dz_1 \dboldq, \\[5pt]
n^3 (\boldx,t) &  = 
\displaystyle \int_{\er^3} p^3(\boldx,z_1,\boldq) \, \dz_1 \dboldq, \\[5pt]
n(\boldx,t) & = n^1(\boldx,t) + n^2(\boldx,t) + n^3(\boldx,t),
\label{totaln}
\end{align}
where $n(\boldx,t)$ is the total density of cells.
Here and hereafter the superscript $i$ denotes a quantity associated
with the $i^{th}$ species, for $ i = 1,2,3$.
If an  evolution equation in $n(\boldx,t)$ alone could be found the problem
would be reduced to the classical case. However we will see that this
is not possible in general. First, we define some additional moments 
that arise in the usual manner from (\ref{r1eq})-(\ref{r3eq}) 
during derivation of moment equations. More precisely, we derive the 
evolution equations for (\ref{momn12}) -- (\ref{totaln}) and for
the following moments 
\begin{align}
\label{momj}
j^i_{v_k} (\boldx,t)  &= 
\int_V \int_{\er^3} v_k p^i(\boldx,\boldv,z_1,\boldq)
\dboldv \dz_1 \dboldq,
\qquad i=1,2; \; k=1,2;\\[5pt]  
n^i_{q_k} (\boldx,t) & = 
\int_V \int_{\er^3} q_k p^i(\boldx,\boldv,z_1,\boldq)
\dboldv \dz_1 \dboldq 
\qquad i=1,2; \; k=1,2;\\[5pt]   
\label{momnq}
n^i_{z} (\boldx,t) & =
\int_V \int_{\er^3} z_1 p^i(\boldx,\boldv,z_1,\boldq)
\dboldv \dz_1 \dboldq
\qquad \qquad i=1,2;\\[5pt]  
\label{momnz}
n^3_{q_k} (\boldx,t) &= 
\int_{\er^3} q_k p^3(\boldx,z_1,\boldq) \dz_1 \dboldq,\\[5pt]
n^3_z (\boldx,t)   & =
\int_{\er^3} z_1 p^3(\boldx,z_1,\boldq) \dz_1 \dboldq.
\label{momn3qz}
\end{align}
Multiplying (\ref{r1eq})-(\ref{r3eq}) by $1$, $v_1$, $v_2$, $q_1$, $q_2$
or $z$ and integrating with respect to $\boldv$, $\boldq$ and $z_1$,
we obtain the evolution equations for moments 
(\ref{momn12}) -- (\ref{momn3qz}). This system of partial differential
equations is not closed - it contains some higher order moments
of the following form
\begin{eqnarray}
m^i_{k_1,k_2,k_3,k_4,k_5} (\boldx,t) 
& = & \hspace*{-5pt}
\int_V \int_{\er^3} 
v_1^{k_1} v_2^{k_2} q_1^{k_3} q_2^{k_4} z_1^{k_5} 
p^i(\boldx,\boldv,z_1,\boldq)
\dboldv \dboldq \dz_1,
\nonumber\\[-5pt]
m^3_{k_1,k_2,k_3} (\boldx,t) 
& = & 
\int_{\er^3} 
q_1^{k_1} q_2^{k_2} z_1^{k_3} 
p^i(\boldx,\boldv,z_1,\boldq)
\dboldq \dz_1, \nonumber
\end{eqnarray}
where $i=1,2$, $k_\alpha$, $\alpha=1,2,3,4,5$, are nonnegative integer, 
and the superscript $k_\alpha$ on terms
in the integral denotes the $k_\alpha$-th power of the corresponding
variable. The simplest way  to close the moment equations is by
setting to zero all higher-order moments which do not appear in
(\ref{momn12}) -- (\ref{momn3qz}). More precisely, we use the
following closure assumption: $ 0 = m^1_{1,1,0,0,0} = m^1_{2,0,0,0,0} =
m^1_{0,2,0,0,0} = m^1_{1,0,1,0,0} = m^1_{0,1,1,0,0} = m^1_{1,0,0,1,0}
= m^1_{0,1,0,1,0} = m^1_{1,0,0,0,1} = m^1_{0,1,0,0,1} =
m^1_{0,0,1,0,1} = m^1_{0,0,0,1,1} = m^1_{0,0,0,0,2} = m^2_{1,0,0,0,1}
= m^2_{0,1,0,0,1} = m^2_{0,0,1,0,1} = m^2_{0,0,0,1,1} =
m^2_{0,0,0,0,2} = m^3_{1,0,1} = m^3_{0,1,1} = m^3_{0,0,2} = 0.  $
This closure assumption can be justified for shallow gradients
of the signal \cite{Erban:2004:ICB}.

Under this assumption, we multiply (\ref{r1eq})-(\ref{r3eq}) by $1$,
$v_1$, $v_2$, $q_1$, $q_2$ or $z$, we integrate with respect to
$\boldv$, $\boldq$ and $z_1$, and we discard the higher-order moments;
the result is the following closed system of 16 macroscopic equations.
\begin{eqnarray}
\dfrac{\partial n^1}{\partial t} 
+  
\dfrac{\partial j^1_{v_1}}{\partial x_1} 
+  
\dfrac{\partial j^1_{v_2}}{\partial x_2} 
&=&
- \lambda_1 n^1
+
\lambda_2 n^2
+
\lambda_3 n^3
+
b_1 n^1_z
+
b_2 n^2_z
+
b_3 n^3_z, 
\label{cequationforn1}\\
\dfrac{\partial n^2}{\partial t} \;
\phantom{+\dfrac{\partial j^1_{v_1}}{\partial x_1}+
\dfrac{\partial j^1_{v_2}}{\partial x_2}}
&=&
\phantom{-}\lambda_1 n^1
-
(\lambda_2 + \lambda_0) n^2
-
b_1 n^1_z
-
b_2 n^2_z, 
\label{cequationforn2}\\
\dfrac{\partial n^3}{\partial t} \;
\phantom{+\dfrac{\partial j^1_{v_1}}{\partial x_1}+
\dfrac{\partial j^1_{v_2}}{\partial x_2}}
&=&
\phantom{-}\lambda_0 n^2
-
\lambda_3 n^3
-
b_3 n^3_z,
\label{cequationforn3}
\end{eqnarray}
%
%%%%%%%%%%%%%%%%%%%%%%%%%%%%%%%%%%%%%%%%%%%%%%%%%%%%%%%%%%%%%%%
%
\begin{eqnarray}
\dfrac{\partial j^1_{v_k}}{\partial t}  -  \dfrac{1}{\tau_d}
\left[\gamma n^1_{q_k} - j^1_{v_k} \right] 
&=& 
- \lambda_1 j^1_{v_k}
+
\lambda_2 j^2_{v_k},
\qquad k = 1,2,           
\label{cequationforjj1} \\
\dfrac{\partial j^2_{v_k}}{\partial t} \phantom{ - \dfrac{1}{\tau_d}\left[\Gamma n^1_{q_k} - j^1_{v_k} \right]}
&= &  
\phantom{-}\lambda_1 j^1_{v_k}
-
(\lambda_2 + \lambda_0) j^2_{v_k},
\qquad k = 1,2,                  
\label{cequationforjj2}
\end{eqnarray}
%
%%%%%%%%%%%%%%%%%%%%%%%%%%%%%%%%%%%%%%%%%%%%%%%%%%%%%%%%%%%%%%%
%
\begin{eqnarray}
\dfrac{\partial n^1_{q_k}}{\partial t}  
- 
\dfrac{1}{\tau_a}
\dfrac{\partial S}{\partial x_k}
n^1
+
\dfrac{1}{\tau_a}
n^1_{q_k}
&=& 
- \lambda_1 n^1_{q_k}
+
\lambda_2 n^2_{q_k}
+
\lambda_3 n^3_{q_k},
\qquad k = 1,2,
\label{cequationfornq1}\\
\dfrac{\partial n^2_{q_k}}{\partial t}  
- 
\dfrac{1}{\tau_a}
\dfrac{\partial S}{\partial x_k}
n^2
+
\dfrac{1}{\tau_a}
n^2_{q_k}
&= &
\phantom{-}\lambda_1 n^1_{q_k}
-
(\lambda_2 + \lambda_0) n^2_{q_k},
\qquad k = 1,2,
\label{cequationfornq2} \\
\dfrac{\partial n^3_{q_k}}{\partial t}  
- 
\dfrac{1}{\tau_a}
\dfrac{\partial S}{\partial x_k}
n^3
+
\dfrac{1}{\tau_a}
n^3_{q_k}
&=& 
\phantom{-}\lambda_0 n^2_{q_k}
-
\lambda_3 n^3_{q_k},
\qquad k = 1,2,
\label{cequationfornq3}
\end{eqnarray}
%
%
%%%%%%%%%%%%%%%%%%%%%%%%%%%%%%%%%%%%%%%%%%%%%%%%%%%%%%%%%%%%%%%
%
%
\begin{eqnarray}
\dfrac{\partial n^1_z}{\partial t}  
- 
\dfrac{\partial S}{\partial t}
n^1
+
\dfrac{1}{\tau_a}
n^1_{z}
- \sum_{i=1}^2\dfrac{\partial S}{\partial x_i} j^1_{v_i}
&= &
- \lambda_1 n^1_{z}
+
\lambda_2 n^2_{z}
+
\lambda_3 n^3_{z},
\label{cequationfornp1}\\
\dfrac{\partial n^2_z}{\partial t}  
- 
\dfrac{\partial S}{\partial t}
n^2
+ 
\dfrac{1}{\tau_a}
n^2_{z} \phantom{- \sum_{i=1}^2\dfrac{\partial S}{\partial x_i}j^1_{i}}
&= &
\phantom{-}\lambda_1 n^1_{z}
-
(\lambda_2 + \lambda_0) n^2_{z},
\label{cequationfornp2} \\
\dfrac{\partial n^3_z}{\partial t}  
- 
\dfrac{\partial S}{\partial t}
n^3
+ 
\dfrac{1}{\tau_a}
n^3_{z}\phantom{- \sum_{i=1}^2\dfrac{\partial S}{\partial x_i}j^1_{i}}
&=& 
\phantom{-}\lambda_0 n^2_{z}
-\lambda_3 n^3_{z}
\label{cequationfornp3}
\end{eqnarray}
Note that the sum of equations 
(\ref{cequationforn1}) -- (\ref{cequationforn3}) is the standard
continuity equation for $n$, \ie,
\begin{equation}
\dfrac{\partial n}{\partial t} 
+  
\dfrac{\partial j^1_{v_1}}{\partial x_1} 
+  
\dfrac{\partial j^1_{v_2}}{\partial x_2} 
= 
0.
\label{contneq}
\end{equation}
The system (\ref{cequationforn1}) -- (\ref{cequationfornp3})
can be written more compactly by defining the following vectors
and matrices. 
\begin{align*}
\boldn =& (n^1,n^2,n^3)^T \qquad \quad 
\boldn_z = (n^1_z,n^2_z,n^3_z)^T  \\[5pt]
\boldn_{\boldq} &= (\boldn_{q_1},\boldn_{q_2}) =(n_{q_1}^1,n_{q_1}^2
n_{q_1}^3,n_{q_2}^1,n_{q_2}^2,n_{q_2}^3)^T  
\end{align*}
\begin{equation*}
\boldj = (\boldj_{v_1},\boldj_{v_2})^T 
= (j_{v_1}^1,j_{v_1}^2,j_{v_2}^1,j_{v_2}^2)^T 
\qquad \qquad 
\nabla 
= 
\left( 
\frac{\partial}{\partial x_1},
\frac{\partial}{\partial x_2}
\right)^T
\end{equation*}
\begin{equation}
% \boldI_3 = \left[\begin{array}{ccc} 
% 1 & 0 & 0 \\
% 0 & 1 & 0 \\
% 0 & 0 & 1 
% \end{array}\right] 
%
\boldLam = \left[\begin{array}{ccc} 
-\lambda_1 & \lambda_2&\lambda_3\\
\lambda_1 & -(\lambda_2+\lambda_0)&0 \\
0 & \lambda_0& - \lambda_3
\end{array}\right] 
\qquad \qquad
\boldLamone = \left[\begin{array}{cc} 
-\lambda_1 & \lambda_2\\
\lambda_1 & -(\lambda_2+\lambda_0) 
\end{array}\right] 
\label{mat1} 
\end{equation}
\begin{equation}
\boldB = \left[\begin{array}{ccc} 
b_1&b_2&b_3\\
-b_1&-b_2&0 \\
0 & 0& -b_3
\end{array}\right] 
\qquad \qquad 
\boldJ =  \left[\begin{array}{cc} 
1&0\\
0&0 \\
0 & 0
\end{array}\right]
\qquad \qquad 
\boldJone =  \left[\begin{array}{cc} 
1&0\\
0&0 
\end{array}\right]
%
% \boldI_2 =  \left[\begin{array}{cc} 
% 1&0\\
% 0&1 
% \end{array}\right]
\label{mat2} 
\end{equation}
We further define the tensor product of an $s_1 \times s_2$ matrix 
$\boldX=\{ x_{ik} \}_{i,k=1}^{s_1,s_2}$ with an $s_3 \times s_4$ matrix 
$\boldY=\{ y_{ik} \}_{i,k=1}^{s_3,s_4}$ to be  the  $(s_1s_3) \times (s_2s_4)$
matrix  
$$
\boldX \otimes \boldY = 
\left[\begin{array}{cccc} 
x_{1,1} \boldY& x_{1,2}\boldY & \dots & x_{1,s_2}\boldY \\
\dots & \dots & \dots  & \dots \\
x_{s_1,1}\boldY & x_{s_1,2} \boldY& \dots & x_{s_1,s_2}\boldY 
\end{array}\right]. 
$$
Then  (\ref{cequationforn1}) -- (\ref{cequationfornp3})
can be rewritten in the form
\begin{eqnarray}
\label{neqn}
\dfrac{\partial \boldn}{\partial t} + \left(\nabla^T \otimes
\boldJ \right)\boldj 
&=& \boldLam \boldn + \boldB \boldn_z, \\
\label{nzeqn}
\dfrac{\partial \boldn_z}{\partial t} 
&=& \dfrac{\partial S}{\partial t} \boldn +\left(\boldLam
-\dfrac{1}{\tau_a}\boldI_3\right)\boldn_z + \left(\nabla^T S \otimes
\boldJ\right)\boldj, \\  
\label{nqeqn}
\dfrac{\partial \boldn_{q_k}}{\partial t}  
&= & \dfrac{(\nabla S)_k}{\tau_a} \boldn + \left(\boldLam
-\dfrac{1}{\tau_a}\boldI_3\right) \boldn_{q_k}  \qquad 
\mbox{ for~~ }  k = 1,2, \quad  \\ 
\label{jeqn}
\dfrac{\partial \boldj_{v_k}}{\partial t}  
&= & \dfrac{\gamma}{\tau_d} \boldJ^T \boldn_{q_k} + \left(
 \boldLamone - \dfrac{1}{\tau_d}\boldJ_1 \right)\boldj_{v_k}  
\qquad
\mbox{ for~~ }  k = 1,2. \quad 
\label{system}
\end{eqnarray}
wherein ${\mathbf I}_k$ is $k \times k$ identity matrix.  This
system can in turn be written more compactly as  the system
\begin{equation} 
\dfrac{\partial \boldU}{\partial t} + \boldD \boldU = \boldA(\boldx,t) \boldU
\label{sysu}
\end{equation}
wherein
\begin{equation}
\boldU = \left(\begin{array}{c} \boldn \\ \boldn_z \\ \boldn_{\boldq}
\\ \boldj
\end{array} \right),
\qquad \quad
\boldD = \left[\begin{array}{cccc} 
{\mathbf 0} \; & \; {\mathbf 0} \; & \; {\mathbf 0} \; &
 {\mathbf \Omega} \\ [5pt] 
{\mathbf 0} \; & \; {\mathbf 0} \; & \; {\mathbf 0} \; & {\mathbf 0} \\ [5pt] 
{\mathbf 0} \; & \; {\mathbf 0} \; & \; {\mathbf 0} \; & {\mathbf 0} \\ [5pt]
{\mathbf 0} \; & \; {\mathbf 0} \; & \; {\mathbf 0} \; & {\mathbf 0}  
\end{array}\right],
\qquad \quad
\mathbf \Omega \equiv \nabla^T \otimes \boldJ,
\label{mata}
\end{equation}
and 
\begin{equation}
\boldA = \left[\begin{array}{cccc}
\boldLam&\boldB&{\mathbf 0}&{\mathbf 0}\\[5pt]
\dfrac{\partial S}{\partial t}\boldI_3 & \left(\boldLam
-\dfrac{1}{\tau_a}\boldI_3\right)&{\mathbf 0}& (\nabla S)^T \otimes \boldJ \\[5pt]
\dfrac{\nabla S}{\tau_a} \otimes \boldI_3&{\mathbf 0}&\boldI_2\otimes\left(\boldLam
-\dfrac{1}{\tau_a}\boldI_3\right)&{\mathbf 0} \\[5pt]
{\mathbf 0}&{\mathbf 0}& \dfrac{\gamma}{\tau_d} \boldI_2 \otimes \boldJ^T &
\boldI_2 \otimes  \left(\boldLamone 
- \dfrac{1}{\tau_d}\boldJone\right) 
\end{array}\right]. \qquad
\label{mata2}
\end{equation} 
We should note that discarding higher-order moments can be justified
for the small signal gradient case in which the second moments of
internal variables are sufficiently small compared to lower- order
moments \cite{Erban:2004:ICB}. It is important to note that the second
order velocity moments $m^1_{1,1,0,0,0},$ $m^1_{2,0,0,0,0}$ and
$m^1_{0,2,0,0,0}$ were also set to zero because we do not have an
obvious moment closure for them similar to what was used in the
bacterial case \cite{Erban:2005:STS}, where the Cattaneo
approximations could be used. 

To obtain a better approximation of these moments we can
follow the reasoning that lead to the closure 
(\ref{jik}) earlier.  To illustrate this, let us modify the taxis model
by adding a random component to the cellular movement, namely we
change the transport equation (\ref{r1eq}) to the following equation
for $p^1$:
\begin{eqnarray}
\cL p^1 
&=& -\nabla_\boldx \cdot v p^1
-  
\nabla_\boldv \cdot \left[
\dfrac{1}{\tau_d}
\left(
\gamma \boldq - \boldv \right) p^1 
\right]
-
\dfrac{\partial}{\partial z_1} 
\left[
\left(
\boldv \cdot \nabla S
\right)
p^1 
\right] \nonumber \\ [8pt]
&& - \lambda p^1   + 
\lambda\int_V T(\boldv,\boldv')
p^1(\boldx,\boldv'\!,t) \dboldv'- (\lambda_1 - b_1 z_1) p^1 \\[8pt]
&&
+ (\lambda_2 + b_2 z_1) p^2  +  \delta_{\boldv} (\lambda_3 + b_3 z_1)
p^3 \nonumber  
 \label{modr1eq}
\end{eqnarray}
where the turning kernel is  given by $T(\boldv,\boldv') =
(2 \pi v_0)^{-1} \delta(|\boldv -\boldv'| - v_0)$
similarly as before, and $v_0$ and $\lambda$
are positive constants. The rationale behind this kernel is to
incorporate some noise into the system, and $v_0$ specifies the
``strength" of this noise. If we follow the previous procedure, we
would obtain the same system of equations (\ref{sysu}), which are
independent of $v_0.$ This is not suprising - we already saw in
Section
\ref{secdirection} that equation (\ref{chemequation}) contains the
diffusion term if the apropriate closure assumption is derived from
(\ref{jik}) and used for the second-order velocity
moments. Similarly as in (\ref{jik}), we can multiply
the transport equations (\ref{modr1eq}) and (\ref{r2eq}) 
by $v_k v_l$, $k,l=1,2$, and
neglect time derivatives of the convective fluxes, third-order
velocity moments and mixed velocity-internal moments to obtain
\begin{eqnarray}
m^1_{2,0,0,0,0} = m^1_{0,2,0,0,0}
&= &
\dfrac{\lambda \tau_d (\lambda_0 +\lambda_2) v_0^2}
{4 (\lambda_0 +\lambda_2) + 2 \tau_d \lambda_0 \lambda_1} 
n^1(\boldx,t), \nonumber \\[-5pt]
\label{secmoms}
\\
m^1_{1,1,0,0,0} &=& 0. \nonumber  
\end{eqnarray}
Using the moment closure (\ref{secmoms}) for the convective momentum 
flux of $p^1$,
we obtain the following  system of moment equations (compare with (\ref{sysu})).
\begin{equation} 
\dfrac{\partial \boldU}{\partial t} + \widetilde{\boldD} \boldU 
= 
\boldA(\boldx,t) \boldU
\label{sysu2}
\end{equation}
Here $\boldA(\boldx,t)$ is given by (\ref{mata2}) and
$\boldU = (\boldn,\boldn_z,\boldn_{\boldq},\boldj)^T$ is as in
(\ref{mata}), but  $\widetilde{\boldD}$ is now given by 
\begin{equation}
\widetilde{\boldD} = \left[\begin{array}{cccc} 
{\mathbf 0} & {\mathbf 0} \; & \; {\mathbf 0} \; & \; {\mathbf \Omega} \\ [5pt] 
{\mathbf 0} & {\mathbf 0} \; & \; {\mathbf 0} \; & \; {\mathbf 0} \\ [5pt] 
{\mathbf 0} & {\mathbf 0} \; & \; {\mathbf 0} \; & \; {\mathbf 0} \\ [5pt]
\alpha {\mathbf \Omega}^T & {\mathbf 0} \; & \; {\mathbf 0} \; & \; {\mathbf 0}  
\end{array}\right]
\qquad
\mbox{where}
\qquad
\alpha = \dfrac{\lambda \tau_d (\lambda_0 +\lambda_2) v_0^2}
{4 (\lambda_0 +\lambda_2) + 2 \tau_d \lambda_0 \lambda_1}.
\label{mataoverD}
\end{equation}

\subsection{Analysis of the statistics of motion} 

\label{secstatmotion}

To illustrate the validity
of reducing the transport equation to the system of hyperbolic
equations, we compute the dependence
of the mean speed of cells on the strength of the underlying
signal.
 
Multiplying equation (\ref{contneq}) by $\boldx$ and
integrating with respect of $\boldx$, we obtain the
equation for the mean speed $\boldv_{av} (t)$
of the cellular population in the following form
\begin{equation}
\boldv_{av} (t)
\equiv
\dfrac{\partial}{\partial t} 
\left[ \dfrac{1}{n_0}
\int_{\er^2} \boldx n(\boldx,t) \dboldx \right]
=
\dfrac{\overline{\boldj^1}}{n_0}
\end{equation}
where 
\begin{equation}
n_0 = \int_{\er^2} n(\boldx,t) \dboldx
\quad
\mbox{and}
\quad
\overline{\boldj^1} = \int_{\er^2} \boldj^1(\boldx,t) \dboldx.
\label{averspeed}
\end{equation}
Here, $n_0$ denotes the total number of cells in the system and
$\overline{\boldj^1}$ is the spatial average of the flux
$\boldj^1=[j^1_{v_1},j^2_{v_2}]$.  Consequently, to estimate the
average speed of cells at a given time we have to estimate
$\overline{\boldj^1}/n_0.$

\begin{figure}
\MpicturesAB{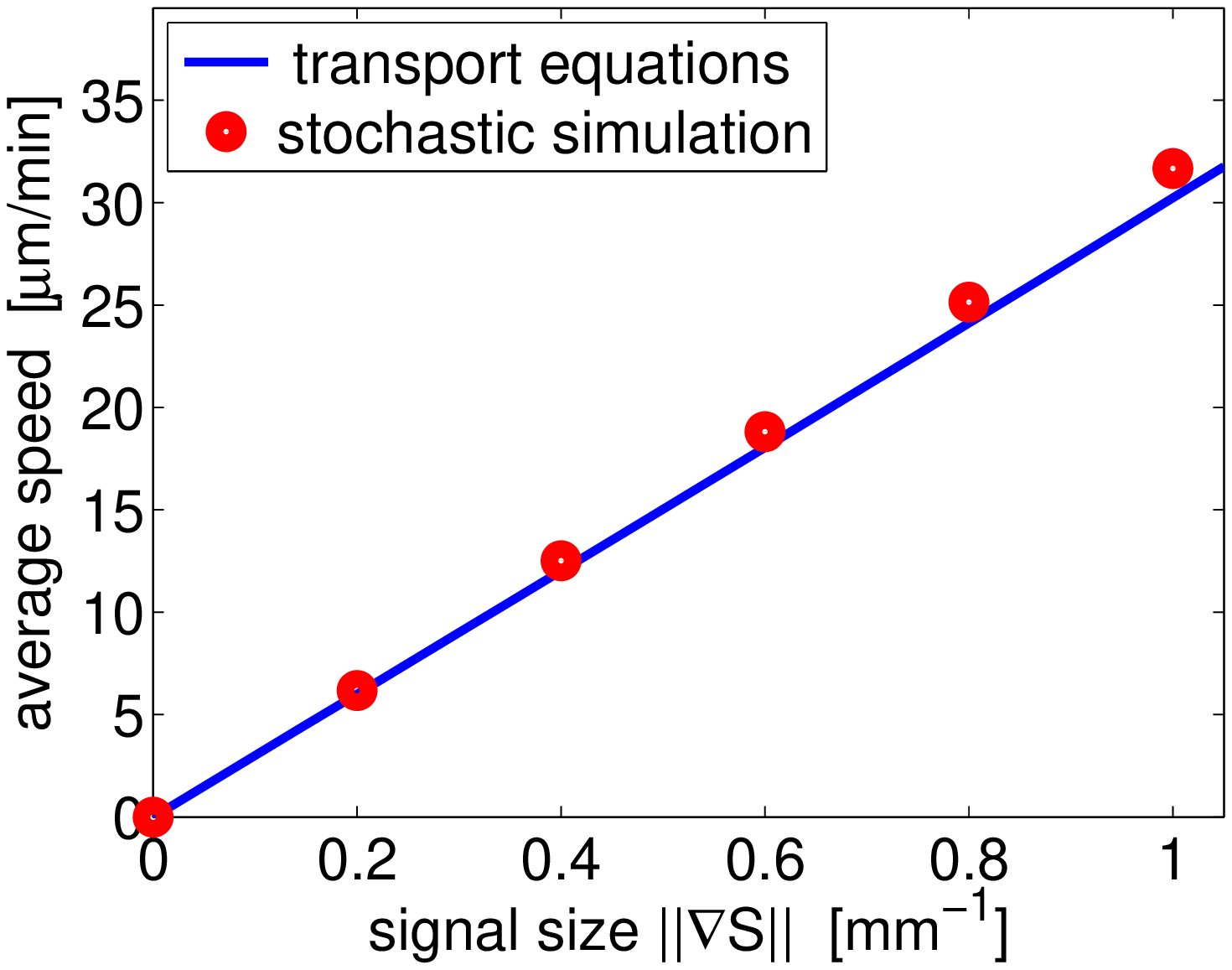}{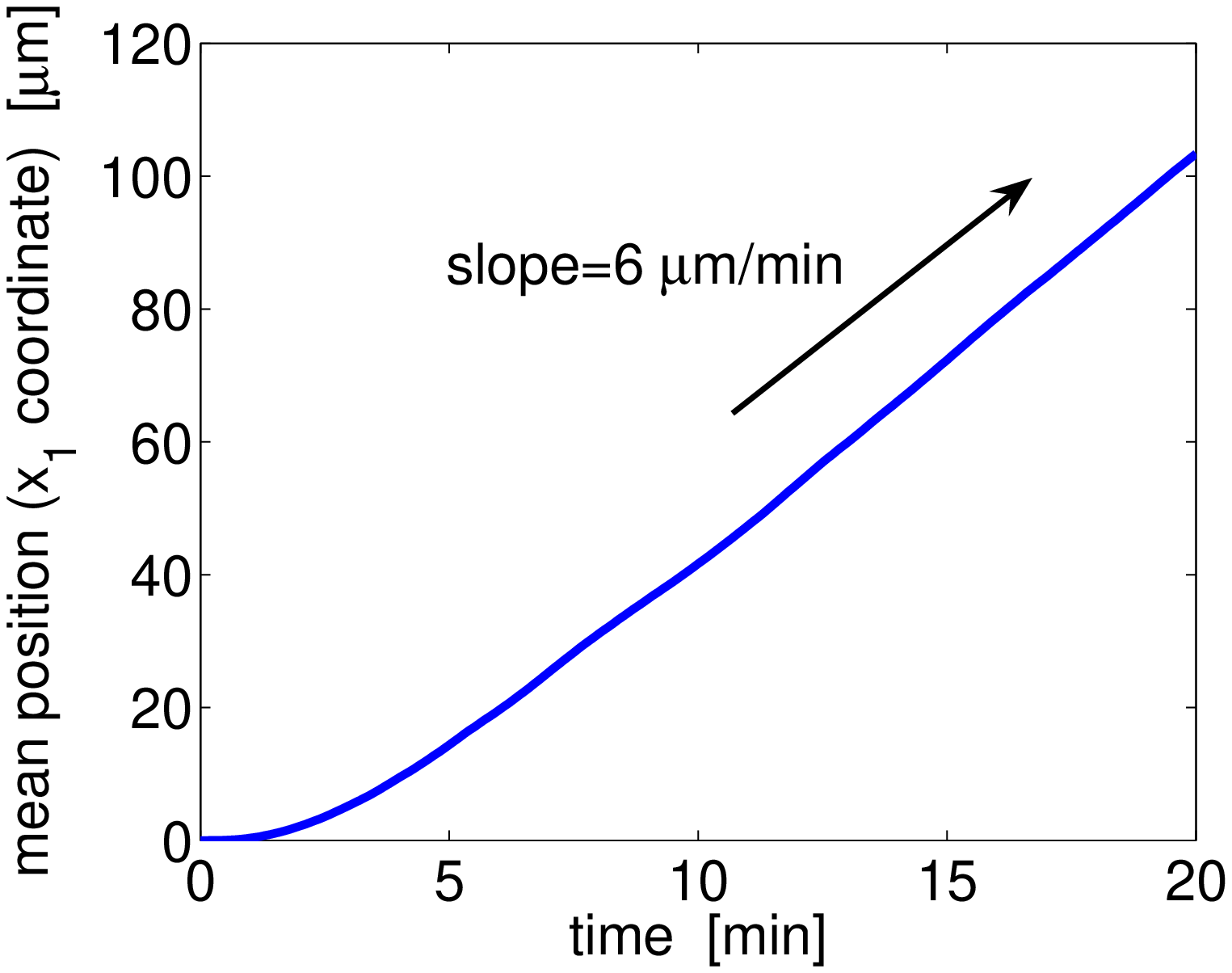}{1.5in}
\caption{(a) {\it Comparison of $\boldv_{av}^{\infty}$ 
computed by $(\ref{theoraverspeed})$ (solid curve) with results
obtained by stochastic simulations (circles) for different
values of $\norm{\!\!\nabla S\!\!}$.}
(b) {\it Average position of individuals (given by stochastic simulation)
as a function of time
for $\norm{\!\!\nabla S\!\!} = 0.2 \; \mbox{\rm mm}^{-1}$.}}
\label{figmepo} 
\end{figure}

To do this we use a one-parameter family of time-independent linear
distributions of extracellular signal $S$ defined by (\ref{testgradient}), 
parametrized by $\norm{\!\! \nabla S \!\!}$. 
Then the matrix $\boldA(\boldx,t) \equiv \boldA(\norm{\!\! \nabla S
\!\!})$ is independent of $\boldx$ and $t$, and  we can integrate
equation (\ref{sysu}) with respect to $\boldx$ to obtain
\begin{equation} 
\dfrac{\partial \overline{\boldU}}{\partial t} 
= \boldA(\norm{\!\! \nabla S \!\!}) \overline{\boldU}
\qquad
\mbox{where}
\qquad
\overline{\boldU} = \int_{\er^2} \boldU(\boldx,t) \dboldx.
\label{sysuover}
\end{equation}
Solving system (\ref{sysuover}) for $\boldU$, we can
estimate the value of mean speed of the cells as
$$
\boldv_{av} (t)
=
\dfrac{\overline{\boldj^1}}{n_0}
=
\dfrac{\overline{U}_{13}}{\overline{U}_{1}+\overline{U}_{2}+\overline{U}_{3}},
$$
and  we see that $\boldv_{av} (t)$ will asymptotically approach
the velocity $\boldv_{av}^{\infty}$ given by 
\begin{equation}
\boldv_{av}^{\infty}
=
\dfrac{\psi_{13}}
{\psi_{1}+ \psi_{2}+ \psi_{3}}
\label{theoraverspeed}
\end{equation}
where $\boldpsi$ is a solution of
\begin{equation*}
\boldA(\norm{\!\! \nabla S \!\!}) \boldpsi = {\mathbf 0}.
\end{equation*}
Using parameter values (\ref{tranparam}) -- (\ref{timeparam})
with $b = 1 \; \mbox{min}^{-1}$ and 
$\gamma = 0.08 \; \mbox{mm}^2$/min,
we can compute the asymptotic average speed $\boldv_{av}^{\infty}$ 
by (\ref{theoraverspeed}) for different values of
$\norm{\!\! \nabla S \!\!}$. The solid curve in Figure \ref{figmepo}(a) 
shows $\boldv_{av}^{\infty}$ as a function of $\norm{\!\! \nabla S \!\!}$.

The theoretical result (\ref{theoraverspeed}) can be verified by
stochastic simulations, and to that end we consider an ensemble of
500 cells. We discretize the boundary (\ref{membrane}) of each cell 
using $m=50$ mesh points. Hence, the state of each cell is   
described by 54-dimensional vector (\ref{discrstate}) similarly as
in Section \ref{secnumsim}. The internal dynamics model $\boldy$ written 
in terms  of (\ref{intradyn1}) is given by 
(\ref{membrane}) -- (\ref{y1eqtheta}).
The radius of a cell is set to $d=7.5\;\mu$m and we use parameter 
values (\ref{tranparam}) -- (\ref{timeparam}) with $b = 1 \; \mbox{min}^{-1}$ 
and $\gamma = 0.08 \; \mbox{mm}^2$/min. The initial conditions
are the same for all cells: namely all cells begin at position
$\boldx = {\mathbf 0}$, their initial velocities satisfy 
$\boldv={\mathbf 0}$, their
internal variables are equal to $0$ around the entire membrane, and
cells are initially unpolarized. The average position of cells 
as a function of time is
given in Figure \ref{figmepo}(b) for $\norm{\!\!\nabla S\!\!} = 0.2 \;
\mbox{\rm mm}^{-1}$.  Since cells are initially unpolarized and
resting, the initial cellular flux is zero. If we wait for a
sufficiently long time, the average speed of the cells relaxes to a
constant, and when we estimate this from long-time simulations, we
obtain the values which are shown as circles in Figure
\ref{figmepo}(a).  Comparing data in Figure \ref{figmepo}(a), 
we see that the theoretical result
(\ref{theoraverspeed}) gives a very good approximation of the mean
asymptotic speed estimated from simulations. This demonstrates the
fact that one can extract population level information from the moment
equations derived earlier.

Finally, we note that the previous analysis can be repeated
for (\ref{sysu2}). The difference between (\ref{sysu}) and
(\ref{sysu2}) is the additional noise in the latter,  which leads
to the system (\ref{sysu2}). However, this noise will only
influence the diffusion constant and the average speed of the 
population will be unchanged. 

\subsection{Further reduction of moment equations}

One can further reduce the size of the system of
moment equations (\ref{sysu}) by supposing that the internal dynamics 
evolve much faster than changes in cytoskeleton
and movement. Considering the excitation
time $\tau_e = 0$ and that the adaptation time $\tau_a \ll \tau_d$,
we can assume the quasi-equilibrium in the equations for
$\boldn_{\boldq}$ and $\boldn_{z}$ in  (\ref{sysu}), \ie,
\begin{equation}
\boldn_{z}
=
\tau_a \left( \boldI_3 - \tau_a \boldLam \right)^{-1}
\left[
\dfrac{\partial S}{\partial t} \, \boldn
+ \left\{
(\nabla S)^T \otimes \boldJ 
\right\}
\,  \boldj
\right]
\label{red1}
\end{equation} 
and
\begin{equation}
\boldn_{\boldq}
=
\left[
\boldI_2\otimes\left(\boldI_3 - \tau_a \boldLam\right)
\right]^{-1}
\left[
\nabla S \otimes \boldn
\right] 
\label{red2}
\end{equation} 
Substituting formulas (\ref{red1}) -- (\ref{red2})
into (\ref{sysu}), we can formally derive the reduced 
system of 7 moment equations for $\boldn$ and $\boldj$ 
only. These equations are derived under the assumption
that $\tau_a$ is small. Passing formally to the limit
$\tau_a \to 0$ in (\ref{red1}) -- (\ref{red2}), 
we obtain $\boldn_{z} \to 0$ and $\boldn_{\boldq} \to
\nabla S \otimes \boldn.$ 
As one would expect in light of the discussion following
(\ref{movvelocity}), the reduced equations predict movement up a steady
gradient, but not in a periodic wave for $\tau_a = 0$.

Another approach to eliminate the internal dynamics
is to assume the quasi-equilibrium assumption
directly in (\ref{bqeq}) -- (\ref{z1eqt0}), i.e.
$$
\boldq(\boldx,t) \approx \nabla S(\boldx,t)
\qquad
\mbox{and}
\qquad
z_1(\boldx,t)
\approx 
\tau_a \dfrac{\partial S}{\partial t}  (\boldx,t)
+ \tau_a \boldv \cdot \nabla S(\boldx,t).
$$
Denoting  $p^1(\boldx,\boldv)$ the density of motile cells at position 
$\boldx \in \er^2$ with velocity $\boldv \in V \subset \er^2$, 
$p^2(\boldx,\boldv)$ the density of resting
polarized cells at position $\boldx$ with polarization axis $\boldv$ and 
$n^3(\boldx)$ the density of resting unpolarized cells,
we can write transport equation for $p^1,$ $p^2$ and $n^3$.
Again 7 moment equations for $\boldn$ and $\boldj$ can be derived.
Such equations were derived and analysed for one-dimensional 
case in \cite{Erban:2005:ICB}. It was shown that in some parameter
regimes, the reduced system approximates the simulation 
with a reasonable precision. See \cite{Erban:2005:ICB} for details.
The precision of the approximation depends on the parameter values
chosen.

\begin{figure}
\MpicturesAB{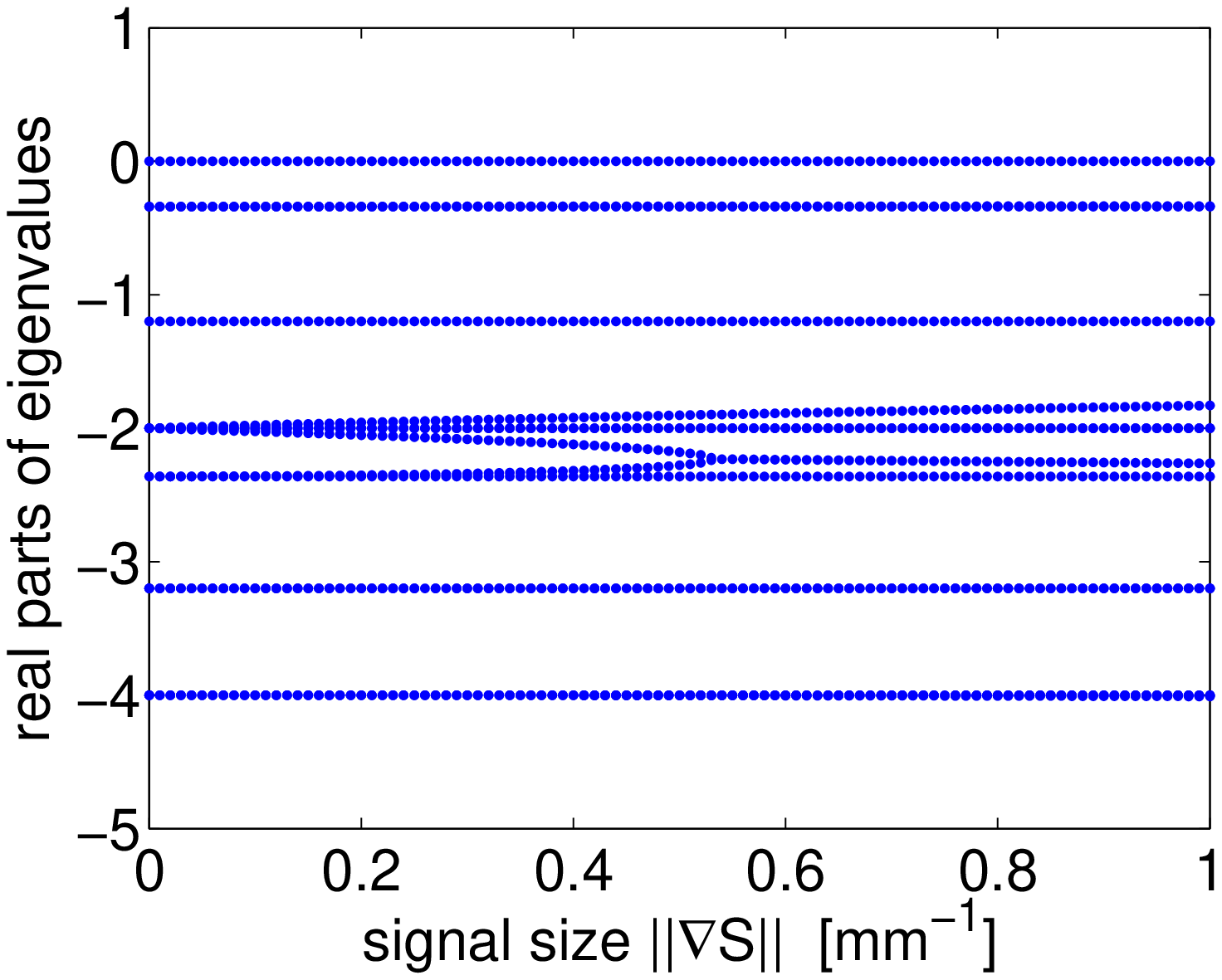}{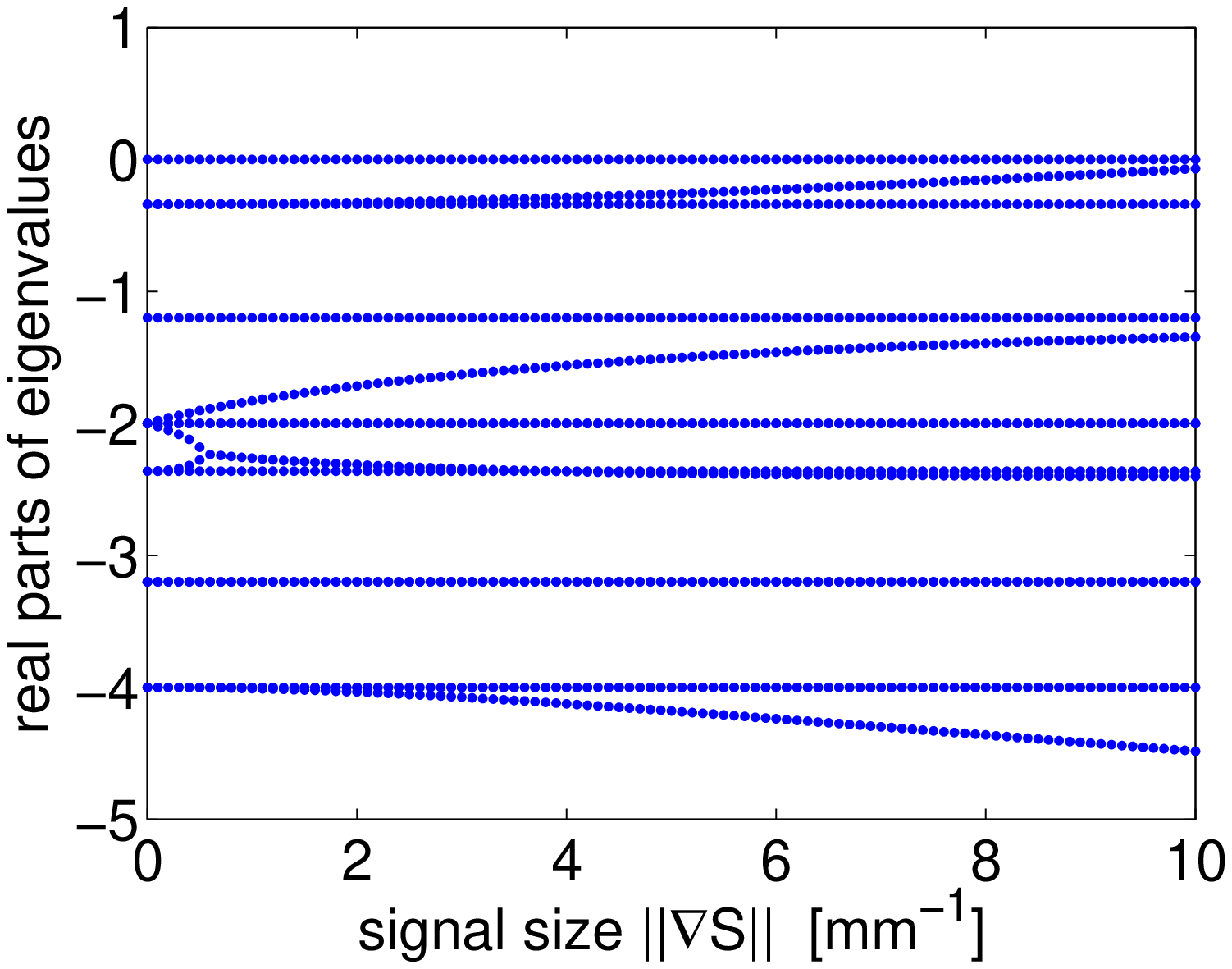}{1.5in}
\caption{{\it The real parts of the  eigenvalues of 
$\boldA(\norm{\!\! \nabla S \!\!})$ 
for}\/ \; (a) $\norm{\!\!\nabla S\!\!} \in [0,1] \; \mbox{mm}^{-1};$
(b) $\norm{\!\!\nabla S\!\!} \in [0,10] \; \mbox{mm}^{-1}.$}
\label{figeig} 
\end{figure}

Finally one can ask whether the method developed in
 \cite{Erban:2004:ICB,Erban:2005:STS}, for reducing a hyperbolic
 system similar to (\ref{sysu}) to the classical chemotaxis equations,
 can be applied here as well. In the context of chemotaxis based on a
 ``run-and-tumble" strategy we were able to analytically compute the
 eigenvalues and eigenvectors of $\boldA(\boldx,t)$, and it was shown
 that they are independent of the chemotactic signal. By exploiting
 the facts that the spectral gaps are signal-independent, and that
 reasonably simple formulas for eigenvectors are available, we reduced
 the hyperbolic system of moment equations to the classical chemotaxis
 description in the bacterial case
 \cite{Erban:2004:ICB,Erban:2005:STS}. In the case of system
 (\ref{sysu}), the resulting slow eigenvalues are, for general values
 of parameters, very complicated functions of the parameters, and in
 particular they depend on the signal. To illustrate this, we consider
 the linear signal distribution  (\ref{testgradient}) that leads to 
 (\ref{sysuover}), we consider parameters from Section \ref{secstatmotion} 
 and plot the real parts of the eigenvalues of $\boldA(\norm{\!\! \nabla S
 \!\!})$ as functions of $\norm{\!\!  \nabla S \!\!}$ in Figure
 \ref{figeig}. One sees there that the eigenvalues vary significantly
 with the signal strength. 

Further analysis is needed to better define the conditions under which
the hyperbolic system can be reduced further. In that vein, we note
that ignoring the term $\sum_{i=1}^2 (\nabla S)_i j^1_{v_i}$
in equation (\ref{cequationfornp1}) leads to the matrix
$\boldA(\norm{\!\! \nabla S \!\!})$ which has signal-independent
eigenvalues. Thus the possibility of further reduction clearly depends
on the closure assumptions.

\section{Discussion and conclusions}

\label{secdiscussion}

The goal in this paper was to derive macroscopic equations for the
collective behavior of amoeboid cells based on models for
individual-level behavior. In previous work we developed a moment
closure approach to the transport equation for a velocity jump process
that describes cell motion for cells that use a ``run-and-tumble''
strategy \cite{Erban:2004:ICB,Erban:2005:STS}.  Here we demonstrated
that this approach can also be applied to the more complex processes
involved in the movement of crawling cells, and showed that one can
predict important macroscopic characteristics from knowledge of the
individual-level properties of these cells. We focused on chemotaxis
in Dd as the model system because much is known about this system, but
the general approach can be applied to any type of extracellular
signal, including those that arise from all receptor-based
interactions of a cell with its environment. Here we summarize the
approach and discuss its advantages and limitations.

In Section \ref{secgensetup} we introduced the general model
(\ref{intradyn1}) -- (\ref{intradyn2}) for the behavior of an
individual eukaryotic cell. Since this model is often
infinite-dimensional, we have to first reduce it to the
finite-dimensional form (\ref{zequation}) -- (\ref{vequation}). If
such a reduction is possible, we can apply the transport equation
framework (\ref{vjpint1}) for the reduced set of variables.  In this
case we can derive the appropriate moment equations (\ref{sysu}) and
use them to study the macroscopic collective properties of cells, as
we illustrated in Section \ref{secstatmotion} where we studied the
dependence of the average speed of the cellular population on the
strength of the extracellular signal.

Therefore the crucial assumption for a model which can be treated
in the framework developed here  is that the projection
${\cal P}$ from (\ref{requirements}) exists and the equations
(\ref{zequation}) -- (\ref{vequation}) can be easily written. If this is
not the case, it may be still possible to reduce the 
individual-level dynamics  to a low-dimensional description
of an individual cell. The behavior of these coarse (intracellular)
observables (on the level of a cell) can be studied by computational 
equation-free methods which are currently being developed
\cite{Kevrekidis:2003:EFM,Erban:2006:GRN}.  The similar 
computational methods can be then used to study the population-level
properties of the amoeboid cells using either the full model of an
individual cell or the best available reduction of it
\cite{Kevrekidis:2003:EFM,Erban:2006:EFC}.

The  moment closure reduction of the transport equation used here can
be justified for small signal gradients  
\cite{Erban:2004:ICB}, but in the case of large signal gradients, the higher 
order moments may not be negligible and cannot be discarded.  Similar
moment methods can be used for any model assuming that the internal
dynamics is close to its quasi-equilibrium. If the original internal
dynamics model is nonlinear, it can be linearized around its
equilibrium value for small signal gradients and the moment approach
can be applied. Hence, the reader should view our linear model as an
example of the linearization of more complex nonlinear problems.  Of
course, the linear model is clearly not valid for large signal
gradients, but this is not the parameter regime studied in this paper.
However,  even some strongly-nonlinear models for internal dynamics
produce simple input-output behavior that can be captured by a linear
model with possibly signal-dependent parameters \cite{Erban:2005:STS},
and thus the results presented here may have broader applicability
than the deriviation would suggest if applied strictly.

In the case of a constant external signal gradient we were able to
derive explicit expressions for various statistics of the motion from
the hyperbolic system derived from the transport equation. We also
discussed the predicted behavior of models for experimental conditions
such as spatio-temporal waves of chemoattractant. It is known that
eukaryotic cells such as Dd or leukocytes aggregate at the source of
the waves, and the models studied here include the processes, such as
adaptation, that  are necessary to reproduce this behavior.

The models described here are all based on deterministic extracellular
signals, although a small random
component was added to the choice of direction. The effects of
stochastic fluctuations in signal detection and processing on movement
were introduced in \cite{Dickinson:1995:TEI}, where stochastic
differential equations are postulated to model cell movement on the
time scale of the molecular processes that govern locomotion.  Some
concrete estimates of the probable role of noise in the signal seen be
a Dd cell were made in
\cite{Othmer:1998:OCS}. However a much more detailed analysis of
stochastic effects in all components of the movement response is
needed.

\begin{acknowledgement}
 Radek Erban's research was supported in part by NSF grant
 DMS 0317372, the Max Planck Institute for Mathematics in the 
 Sciences, the Minnesota Supercomputing Institute,
 Biotechnology and Biological Sciences Research Council, 
 University of Oxford, University of Minnesota and 
 Linacre College, Oxford. Hans G. Othmer's research was
 supported in part by NIH grant GM 29123, NSF grants DMS 9805494
 and  DMS 0317372, the Max Planck Institute for Mathematics in the 
 Sciences and the Alexander von Humboldt Foundation.
\end{acknowledgement}

\bibliographystyle{amsplain}
\bibliography{bibrad}
\end{document}